\documentclass[aps,prd,preprint,tightenlines,superscriptaddress,showpacs,byrevtex]{revtex4}
\usepackage{graphicx}% Include figure files
\usepackage{dcolumn}% Align table columns on decimal point
\usepackage{bm}% bold math

\begin{document}

\vspace*{-3\baselineskip}
\resizebox{!}{3cm}{\includegraphics{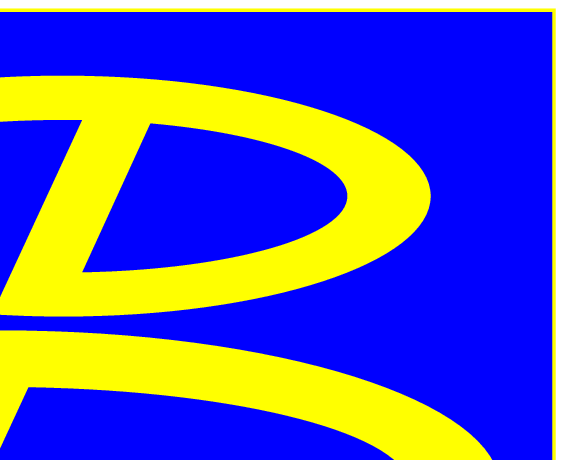}}

%%%%%%%%%%%%%%%%%%%%%%%%%%%%%%%%%%%%%%%%%%%%%%%
\preprint{\tighten\vbox{\hbox{\hfil KEK preprint 2002-131}
                        \hbox{\hfil Belle preprint 2003-1 }
}}

%%%%%%%%%%%%%%%%%%%%%%
% New commands
%%%%%%%%%%%%%%%%%%%%%%
\newcommand{\atoyerr}{\pm 0.27}
\newcommand{\stoyerr}{\pm 0.41}
\newcommand{\svalue}{-1.23}
\newcommand{\svaluecnst}{-0.82}
\newcommand{\sstaterr}{^{+0.24}_{-0.15}}
\newcommand{\ssyserr}{^{+0.08}_{-0.07}}
\newcommand{\sresult}{\svalue~\stoyerr({\rm stat})~\ssyserr({\rm syst})}
\newcommand{\cvalue}{+0.77}
\newcommand{\cvaluecnst}{+0.57}
\newcommand{\cstaterr}{^{+0.20}_{-0.23}}
\newcommand{\csyserr}{\pm 0.08}
\newcommand{\cresult}{\cvalue~\atoyerr({\rm stat})~\csyserr({\rm syst})}
\newcommand{\avalue}{\cvalue}
\newcommand{\avaluecnst}{\cvaluecnst}
\newcommand{\astaterr}{\cstaterr}
\newcommand{\asyserr}{\csyserr}
\newcommand{\aresult}{\cresult}
\newcommand{\rt}{\rightarrow}
\newcommand{\pipi}{\pi\pi}
\newcommand{\spipi}{{\cal S}_{\pipi}}
\newcommand{\apipi}{{\cal A}_{\pipi}}
\newcommand{\rpipi}{{\cal R}}
\newcommand{\skpi}{{\cal S}_{K\pi}}
\newcommand{\akpi}{{\cal A}_{K\pi}}
\newcommand{\taub}{{\tau}_{B^0}}
\newcommand{\bz}{B^0}
\newcommand{\bzb}{\overline{B}{}^0}
\newcommand{\dmd}{\Delta m_d}
\newcommand{\cldd}{99.93\%}
\newcommand{\cla}{0.993}
\newcommand{\cls}{0.997}
\newcommand{\MBC}{M_{\rm bc}}

\newcommand{\apipif}{{\cal A}_{\pipi}}
\newcommand{\spipif}{{\cal S}_{\pipi}}
\newcommand{\akpif}{{\cal A}_{K\pi}}
\newcommand{\skpif}{{\cal S}_{K\pi}}
\newcommand{\abkgf}{{\cal A}_{\rm bkg}}
\newcommand{\dspipif}{\Delta \spipif}
\newcommand{\xapipi}{x_{{\cal A}\pi\pi}}
\newcommand{\xspipi}{x_{{\cal S}\pi\pi}}
\newcommand{\xamin}{x_{{\cal A}1}}
\newcommand{\xamax}{x_{{\cal A}2}}
\newcommand{\abest}{{\cal A}_{\rm best}}

%
%\vspace {1cm}
%
\title{  \Large Evidence for
     {\boldmath $CP$}-Violating Asymmetries 
     in $B^0 \to \pi^+\pi^-$ Decays
    and Constraints on the CKM Angle {\boldmath $\phi_2$}}

%\input author
%%%%% Force institutions to appear in alphabetical order when typeset.
\affiliation{Aomori University, Aomori}
\affiliation{Budker Institute of Nuclear Physics, Novosibirsk}
%%%\affiliation{Chiba University, Chiba}
\affiliation{Chuo University, Tokyo}
\affiliation{University of Cincinnati, Cincinnati, Ohio 45221}
\affiliation{University of Frankfurt, Frankfurt}
%%%\affiliation{Gyeongsang National University, Chinju}
\affiliation{University of Hawaii, Honolulu, Hawaii 96822}
\affiliation{High Energy Accelerator Research Organization (KEK), Tsukuba}
\affiliation{Hiroshima Institute of Technology, Hiroshima}
\affiliation{Institute of High Energy Physics, Chinese Academy of Sciences, Beijing}
\affiliation{Institute of High Energy Physics, Vienna}
\affiliation{Institute for Theoretical and Experimental Physics, Moscow}
\affiliation{J. Stefan Institute, Ljubljana}
\affiliation{Kanagawa University, Yokohama}
\affiliation{Korea University, Seoul}
\affiliation{Kyoto University, Kyoto}
\affiliation{Kyungpook National University, Taegu}
\affiliation{Institut de Physique des Hautes \'Energies, Universit\'e de Lausanne, Lausanne}
\affiliation{University of Ljubljana, Ljubljana}
\affiliation{University of Maribor, Maribor}
\affiliation{University of Melbourne, Victoria}
\affiliation{Nagoya University, Nagoya}
\affiliation{Nara Women's University, Nara}
\affiliation{National Kaohsiung Normal University, Kaohsiung}
\affiliation{National Lien-Ho Institute of Technology, Miao Li}
\affiliation{National Taiwan University, Taipei}
\affiliation{H. Niewodniczanski Institute of Nuclear Physics, Krakow}
\affiliation{Nihon Dental College, Niigata}
\affiliation{Niigata University, Niigata}
\affiliation{Osaka City University, Osaka}
\affiliation{Osaka University, Osaka}
\affiliation{Panjab University, Chandigarh}
\affiliation{Peking University, Beijing}
\affiliation{Princeton University, Princeton, New Jersey 08545}
\affiliation{RIKEN BNL Research Center, Upton, New York 11973}
\affiliation{Saga University, Saga}
\affiliation{University of Science and Technology of China, Hefei}
\affiliation{Seoul National University, Seoul}
\affiliation{Sungkyunkwan University, Suwon}
\affiliation{University of Sydney, Sydney NSW}
%%%\affiliation{Tata Institute of Fundamental Research, Bombay}
\affiliation{Toho University, Funabashi}
\affiliation{Tohoku Gakuin University, Tagajo}
\affiliation{Tohoku University, Sendai}
\affiliation{University of Tokyo, Tokyo}
\affiliation{Tokyo Institute of Technology, Tokyo}
\affiliation{Tokyo Metropolitan University, Tokyo}
\affiliation{Tokyo University of Agriculture and Technology, Tokyo}
%%%\affiliation{Toyama National College of Maritime Technology, Toyama}
\affiliation{University of Tsukuba, Tsukuba}
\affiliation{Utkal University, Bhubaneswer}
\affiliation{Virginia Polytechnic Institute and State University, Blacksburg, Virginia 24061}
\affiliation{Yokkaichi University, Yokkaichi}
\affiliation{Yonsei University, Seoul}
  \author{K.~Abe}\affiliation{High Energy Accelerator Research Organization (KEK), Tsukuba} % KEK
  \author{K.~Abe}\affiliation{Tohoku Gakuin University, Tagajo} % TohokuGakuin
  \author{N.~Abe}\affiliation{Tokyo Institute of Technology, Tokyo} % TIT
% \author{R.~Abe}\affiliation{Niigata University, Niigata} % Niigata
  \author{T.~Abe}\affiliation{Tohoku University, Sendai} % Tohoku
  \author{I.~Adachi}\affiliation{High Energy Accelerator Research Organization (KEK), Tsukuba} % KEK
% \author{Byoung~Sup~Ahn}\affiliation{Korea University, Seoul} % Korea
  \author{H.~Aihara}\affiliation{University of Tokyo, Tokyo} % Tokyo
  \author{K.~Akai}\affiliation{High Energy Accelerator Research Organization (KEK), Tsukuba} % KEK
  \author{M.~Akatsu}\affiliation{Nagoya University, Nagoya} % Nagoya
  \author{M.~Akemoto}\affiliation{High Energy Accelerator Research Organization (KEK), Tsukuba} % KEK
% \author{M.~Asai}\affiliation{Hiroshima Institute of Technology, Hiroshima} % Hiroshima
  \author{Y.~Asano}\affiliation{University of Tsukuba, Tsukuba} % Tsukuba
% \author{T.~Aso}\affiliation{Toyama National College of Maritime Technology, Toyama} % Toyama
% \author{V.~Aulchenko}\affiliation{Budker Institute of Nuclear Physics, Novosibirsk} % BINP
% ...
% Ljubljana
% \author{D.~Z\"urcher}\affiliation{Institut de Physique des Hautes \'Energies, Universit\'e de Lausanne, Lausanne} % Lausanne
\collaboration{The Belle Collaboration}

\date{\today}% It is always \today, today,

\begin{abstract}
We present an improved measurement of
$CP$-violating asymmetries in $B^0 \rightarrow \pi^+\pi^-$ decays
based on
a $78~{\rm fb}^{-1}$ data sample collected at the $\Upsilon(4S)$ resonance
with the Belle detector at the KEKB asymmetric-energy $e^+e^-$ collider.
We reconstruct one neutral $B$ meson
as a  $B^0 \rightarrow \pi^+\pi^-$
$CP$ eigenstate and identify
the flavor of the accompanying $B$ meson
from inclusive properties of its decay products.
We apply an unbinned maximum likelihood fit to the
distribution of the time intervals between the two $B$ meson decay points.
The fit yields the $CP$-violating asymmetry amplitudes
$\apipif = \aresult$
and
$\spipif = \sresult$,
where the statistical uncertainties are determined from Monte Carlo
pseudo-experiments.
We obtain confidence intervals
for $CP$-violating asymmetry parameters $\apipi$ and $\spipi$
based on a frequentist approach.
We rule out the $CP$-conserving case, $\apipi=\spipi=0$, at 
the $\cldd$ confidence level.
We discuss how these results constrain the 
value of the CKM angle $\phi_2$.

\end{abstract}

%%% \pacs{PACS numbers: 11.30.Er, 12.15.Hh, 13.25.Hw}

\maketitle

\section{INTRODUCTION}
In 1973, Kobayashi and Maskawa (KM) proposed a model where $CP$ violation is
accommodated as an irreducible complex phase in the
weak-interaction quark mixing matrix~\cite{KM}.
Recent measurements of the $CP$-violating parameter $\sin 2\phi_1$ 
by the Belle~\cite{CP1_Belle,CP1_Belle3} and BaBar~\cite{CP1_BaBar} collaborations 
established $CP$ violation in the neutral $B$ meson system that
is consistent with the KM model.
Measurements of other $CP$-violating parameters provide important
tests of the KM model. 

The KM model predicts $CP$-violating asymmetries in the time-dependent
rates for  $B^0$ and $\bzb$
decays to a common $CP$ eigenstate, $f_{CP}$~\cite{Sanda}.
In the decay chain $\Upsilon(4S)\to \bz\bzb \to f_{CP}f_{\rm tag}$,
in which one of the $B$ mesons decays at time $t_{CP}$ to $f_{CP}$ 
and the other decays at time $t_{\rm tag}$ to a final state
$f_{\rm tag}$ that distinguishes between $B^0$ and $\bzb$, 
the decay rate has a time dependence
given by~\cite{CPVrev}
\begin{eqnarray}
\label{eq:R_q}
{\cal P}_{\pi\pi}^q(\Delta{t}) = 
\frac{e^{-|\Delta{t}|/{\taub}}}{4{\taub}}
\left[1 + q\cdot 
\left\{ \spipi\sin(\dmd\Delta{t})   \right. \right. \nonumber \\
\left. \left.
   + \apipi\cos(\dmd\Delta{t})
\right\}
\right],
\end{eqnarray}
where $\taub$ is the $B^0$ lifetime, $\dmd$ is the mass difference 
between the two $B^0$ mass
eigenstates, $\Delta{t}$ = $t_{CP}$ $-$ $t_{\rm tag}$, and
the $b$-flavor charge $q$ = +1 ($-1$) when the tagging $B$ meson
is a $B^0$ ($\bzb$).
The $CP$-violating parameters $\spipi$ and $\apipi$ 
defined in Eq.~(\ref{eq:R_q}) are expressed as
\begin{eqnarray}
\apipi = \frac{|\lambda|^2 - 1}{|\lambda|^2 + 1},~~~
\spipi = \frac{2Im \lambda}{|\lambda|^2 + 1},
\end{eqnarray}
where $\lambda$ is a complex 
parameter that depends on both $\bz$-$\bzb$
mixing and on the amplitudes for $\bz$ and $\bzb$ decay to 
$\pi^+\pi^-$. In the Standard Model, to a good approximation,
$|\lambda|$ is equal to the absolute value
of the ratio of the $\bzb$ to $\bz$ decay amplitudes.
A measurement of time-dependent $CP$-violating asymmetries
in the mode $\bz \to \pi^+\pi^-$~\cite{CC} 
is sensitive to direct $CP$ violation and the CKM angle $\phi_2$~\cite{alpha}.
If the decay proceeded only via a $b\rt u$ tree
amplitude, we would have
$\spipi = \sin 2\phi_2$ and
$\apipi =0$, or equivalently $|\lambda| = 1$.
The situation is complicated by the possibility 
of significant contributions from gluonic $b\to d$
penguin amplitudes that
have a different weak phase and additional strong 
phases~\cite{pipipenguin}.
As a result, 
$\spipi$ may not be equal to $\sin2\phi_2$ and
direct $CP$ violation, $\apipi \neq 0$,  may occur.

 Belle's earlier published study~\cite{Acp_pipi_Belle} 
was based on a 41.8~fb$^{-1}$
data sample containing 44.8 $\times$ 10$^6$ $B\overline{B}$ pairs produced at the $\Upsilon(4S)$ resonance. 
The result suggested large direct $CP$ asymmetry and/or mixing-induced asymmetry in 
$B^0 \rightarrow \pi^+\pi^-$ decay while the corresponding result by the BaBar collaboration
%%v4<<
based on a sample of 88 $\times$ 10$^6$ $B\overline{B}$ pairs
%%v4>>
did not~\cite{apipi_BaBar}.
In this paper,
we report an updated measurement that is based on a $78~{\rm fb}^{-1}$
data sample, containing 85$\times$10$^6$ $B\overline{B}$ pairs.
The most important change is that we determine the statistical significance 
and uncertainties in the $CP$ parameters from the distributions of 
the results of fits to Monte Carlo (MC) pseudo-experiments,
instead of from errors obtained by the likelihood fit to experimental data.
In addition, we have made three significant improvements to the analysis:
a new track reconstruction algorithm that provides better 
performance; a new proper-time interval resolution function that reduces 
systematic uncertainties; and the inclusion of additional signal
candidates by optimizing the cuts for continuum background suppression.

In Sec.~\ref{sec:KEKBBELLE} we describe the KEKB collider and 
the Belle detector.
The reconstruction of $B^0 \rightarrow \pi^+\pi^-$ decays is described 
in Sec.~\ref{sec:recb}.  
The flavor-tagging procedure and vertex reconstruction are
described in Secs.~\ref{sec:FLTAG} and \ref{sec:vertex}. 
After discussing the signal yield in Sec.~\ref{sec:yield}
and introducing the method to
measure $\apipi$ and $\spipi$ from $\Delta t$ distributions in
Sec.~\ref{sec:MLH},
we present the results of the fit in Sec.~\ref{sec:result}, and 
discuss constraints on $\phi_2$ in Sec.~\ref{sec:phi2}.
We summarize the results in Sec.~\ref{sec:conclusion}.

\section{EXPERIMENTAL APPARATUS}
\label{sec:KEKBBELLE}

The data reported here were collected with
the Belle detector at the KEKB asymmetric-energy
$e^+e^-$ collider~\cite{KEKB}, which collides 8.0 GeV $e^-$ 
and 3.5 GeV $e^+$
beams at a small ($\pm$11 mrad) crossing angle.
The $\Upsilon(4S)$ is produced
with a Lorentz boost of $\beta\gamma=0.425$ nearly along
the electron beamline ($z$).
Since the $B^0$ and $\bzb$ mesons are approximately at 
rest in the $\Upsilon(4S)$ center-of-mass system (cms),
$\Delta t$ can be determined from $\Delta z$,
the displacement in $z$ between the $f_{CP}$ and $f_{\rm tag}$ decay vertices:
$\Delta t \simeq (z_{CP} - z_{\rm tag})/\beta\gamma c
 \equiv \Delta z/\beta\gamma c$.

The Belle detector is a large-solid-angle
general purpose spectrometer that
consists of a silicon vertex detector (SVD),
a central drift chamber (CDC), an array of
aerogel threshold \v{C}erenkov counters (ACC), 
time-of-flight scintillation counters, and an electromagnetic calorimeter
comprised of CsI(Tl) crystals  located inside 
a superconducting solenoid coil that provides a 1.5~T
magnetic field.  An iron flux return located outside of
the coil is instrumented to detect $K_L^0$ mesons and muons.  
For more details, see Ref.~\cite{Belle}.

\section{RECONSTRUCTION OF $B^0 \rightarrow \pi^+\pi^-$ DECAYS}
\label{sec:recb}

The $B^0 \to \pi^+\pi^-$ event selection is described
in detail elsewhere~\cite{pipi}. We use oppositely charged track pairs
that are positively identified as pions according to the
likelihood ratio for a particle to be a $K^\pm$ meson, 
KID = $\cal L$($K$)/[$\cal L$($K$)+$\cal L$($\pi$)], which is based on 
the combined information from the ACC and the CDC $dE/dx$ measurements.
Here we use KID$<$0.4 as the default requirement
for the selection of pions.
%%v6<<
For tracks in the momentum range that covers the $B^0\rightarrow\pi^+\pi^-$ 
signal,
%%v6 This requirement has a pion efficiency of 91$\%$ and a misidentification 
%%v6 rate from kaons 
this requirement has a pion efficiency of 91$\%$ and 
10.3$\%$ of kaons are misidentified as pions
%%v4<<
%%v4 of 10$\%$ for the $B^0\rightarrow\pi^+\pi^-$ tracks.
( 10.0$\pm 0.2\%$ from $K^-$ and 10.6$\pm 0.2\%$ from $K^+$ ).
%%v6 for tracks in the momentum range for
%%v6 the $B^0\rightarrow\pi^+\pi^-$ signal region, 
%%v6 which results in the relative charge asymmetry in a misidentification
%%v6 from kaons of $-3\%$.
%%v4>>

 We also select $B^0 \rightarrow K^+\pi^-$ candidates,
which have the same track topology as 
$B^0 \rightarrow \pi^+\pi^-$ candidates, by positively
identifying one charged track as a kaon and the other as a pion.
We use KID$>$0.6 for the
selection of kaons. This requirement has a kaon efficiency of 84$\%$ and
a misidentification rate from pions of 5$\%$.

Candidate $B$ mesons are reconstructed using
the energy difference 
$\Delta E\equiv E_B^{\rm cms} - E_{\rm beam}^{\rm cms}$
and the beam-energy constrained
mass $M_{\rm bc}\equiv\sqrt{(E_{\rm beam}^{\rm cms})^2-(p_B^{\rm cms})^2}$,
where $E_{\rm beam}^{\rm cms}$ is the cms beam energy,
and $E_B^{\rm cms}$ and $p_B^{\rm cms}$ are the cms energy and momentum
of the $B$ candidate.
The signal region is defined as 
$5.271 ~{\rm GeV/c^2}<\MBC<5.287 ~{\rm GeV/c^2}$
and $|\Delta{E}|<0.057$ ~GeV, corresponding to $\pm{3}\sigma$ from
the central values.
In order to suppress background from the $e^+e^- \rightarrow q\overline{q}$
continuum ($q = u,~d,~s,~c$),  we form signal and background
likelihood functions, ${\cal L}_S$ and ${\cal L}_{BG}$, 
from two variables. One is a Fisher
discriminant determined from six modified Fox-Wolfram
moments~\cite{SFW};
the other is the cms $B$ flight direction
with respect to the $z$ axis.
We determine ${\cal L}_S$ from a GEANT-based Monte Carlo (MC) simulation~\cite{bib:Geant}, 
and ${\cal L}_{BG}$ from sideband data in the $5.20~{\rm GeV/c^2}<\MBC<5.26   
~{\rm GeV/c^2}$ and $-0.3~{\rm GeV}<\Delta{E}<0.5$ ~GeV region. We
reduce the continuum background by 
imposing requirements on the likelihood ratio
$LR$ =  ${\cal L}_S/({\cal L}_S+{\cal L}_{BG})$
for candidate events, as described below.

\section{FLAVOR TAGGING}
\label{sec:FLTAG}
Leptons, kaons, and charged pions
that are not associated with the reconstructed
$B^0 \rightarrow \pi^+\pi^-$ decay are used to identify
the flavor of the accompanying $B$ meson.
We apply the same method used for the Belle $\sin 2\phi_1$ 
measurement~\cite{CP1_Belle3}.
We use two parameters, $q$ and $r$, to represent the tagging information.
The first, $q$, is 
defined in Eq.~(\ref{eq:R_q}).
The parameter $r$ is an event-by-event,
MC-determined flavor-tagging dilution factor 
that ranges from $r=0$ for no flavor
discrimination to $r=1$ for {unambiguous flavor assignment.
It is used only to sort data into six $r$ intervals.
%(boundaries at 0.25, 0.5, 0.625, 0.75 and 0.875).
The wrong tag fractions for the six $r$ intervals, $w_l\ (l=1,6)$,
are determined 
from the data and are summarized in Table~\ref{tbl:wtfrac}.
%%%%%%%%%%%%%%%%%%%%%%%%%%%%%%%%%%%%%%%%%%%%%
\begin{table}[!htb]
\caption{The wrong tag fraction $w_l$ for each $r$ interval. 
The errors include both statistical and systematic uncertainties.}
\label{tbl:wtfrac}
\begin{ruledtabular}
\begin{tabular}{ccc}
$l$ & $r$ interval & $w_l$ \\ \hline
1 & 0.000 $-$ 0.250 & 0.458$\pm$0.007 \\
2 & 0.250 $-$ 0.500 & 0.336$\pm$0.010 \\
3 & 0.500 $-$ 0.625 & 0.228$\pm$0.011 \\
4 & 0.625 $-$ 0.750 & 0.160$\pm{0.014}$ \\
5 & 0.750 $-$ 0.875 & 0.112$\pm$0.015 \\
6 & 0.875 $-$ 1.000 & 0.020$\pm$0.007 \\
\end{tabular}
\end{ruledtabular}
\end{table}
%%%%%%%%%%%%%%%%%%%%%%%%%%%%%%%%%%%%%%%%%%%%%

%%v4<<
%%v4 In the previous publication~\cite{Acp_pipi_Belle}, we required $LR$ $>$ 0.825
%%v4 for each flavor tagging interval $r$ in order to suppress the continuum 
%%v4 background.
%%v4 In this analysis, we optimize the expected sensitivity by including
%%v4 additional events with $LR<0.825$ for each $r$ interval.
%%v4 This results in twelve distinct regions ($m$ = 1, 12) in the $LR$-$r$ plane.
%%v4 The $LR$ requirements vary for different regions as shown in
%%v4 Table~\ref{tbl:frac}.
In the previous publication~\cite{Acp_pipi_Belle}, 
we required $LR$ $>$ 0.825 for all candidate events,
while in this analysis we optimize the expected sensitivity by including
additional candidate events with a lower signal likelihood ratio.
%%v6<<
%%v6 The requirements on $LR$ vary for different tagging dilution factor $r$
The requirements on $LR$ vary for different values of tagging dilution factor $r$,
as indicated in Table~\ref{tbl:frac},
since the separation of continuum background from the $B$ signal varies with $r$;
there are 12 distinct regions in the $LR$-$r$ plane.
%%v4>>

\section{Vertex reconstruction}
\label{sec:vertex}
The vertex reconstruction algorithm is the same as 
that used for the sin2$\phi_1$ analysis~\cite{CP1_Belle3}.
The vertex positions for the $f_{CP}$ decay ($\pi^+\pi^-$) 
and $f_{\rm tag}$ decays are reconstructed using tracks 
with associated hits in the SVD.
Each vertex position is also constrained by
the interaction point profile, determined run-by-run, smeared in the
$r$-$\phi$ plane to account for the $B$ meson decay length.
With these requirements, we are able to determine a vertex even with a single
track; the fraction of single-track vertices is about 10$\%$ for $z_{CP}$ and
22$\%$ for $z_{\rm tag}$.
The $f_{\rm tag}$ vertex is determined from all 
well-reconstructed tracks, 
excluding the $B^0 \to \pi^+\pi^-$ tracks and
tracks that form a $K_S^0$ candidate.

%%%%%%%%%%%%%%%%%%%%%%%%
\begin{figure}[!htb]
\begin{center}
\resizebox{0.4\textwidth}{!}{\includegraphics{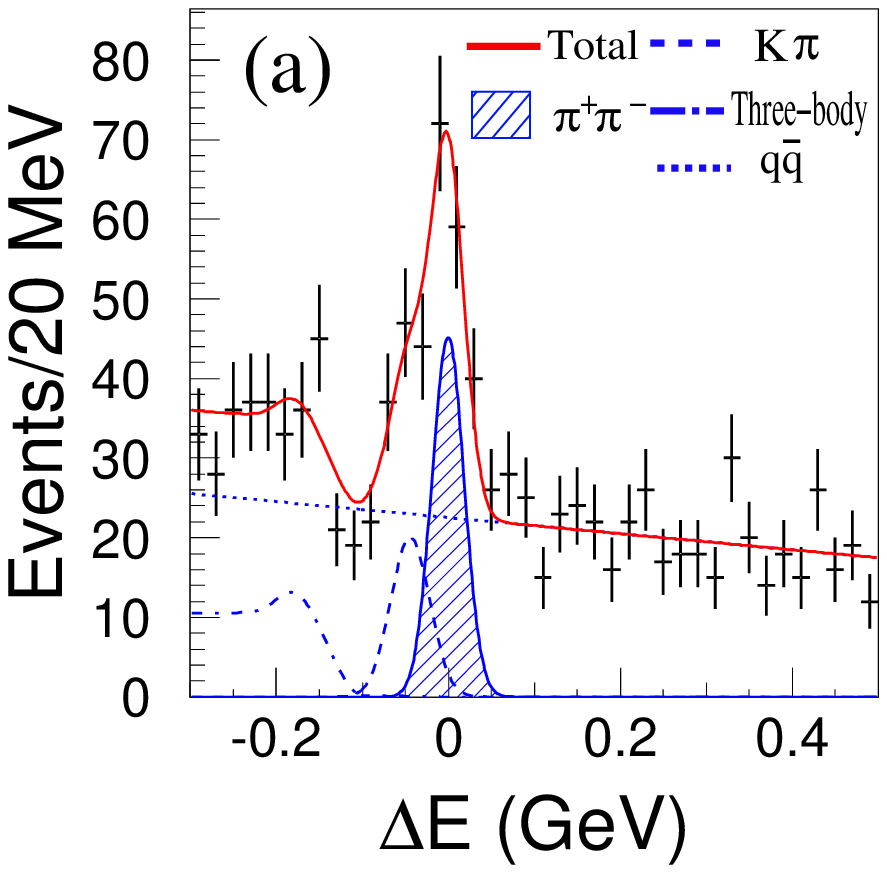}}
\resizebox{0.4\textwidth}{!}{\includegraphics{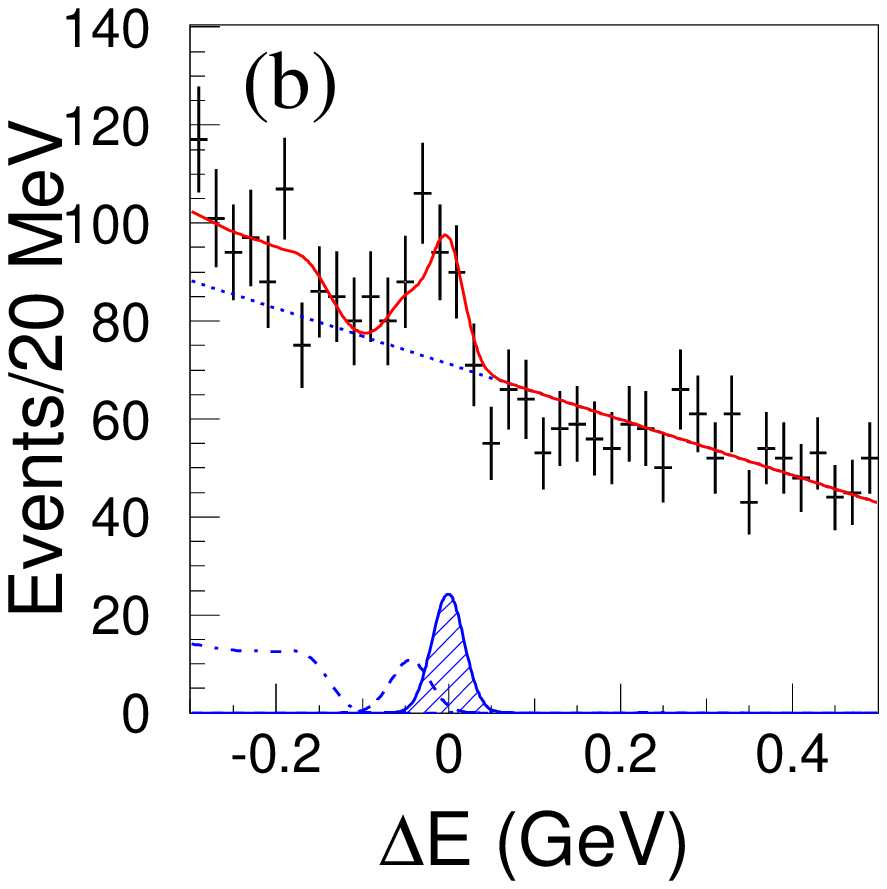}}
\resizebox{0.4\textwidth}{!}{\includegraphics{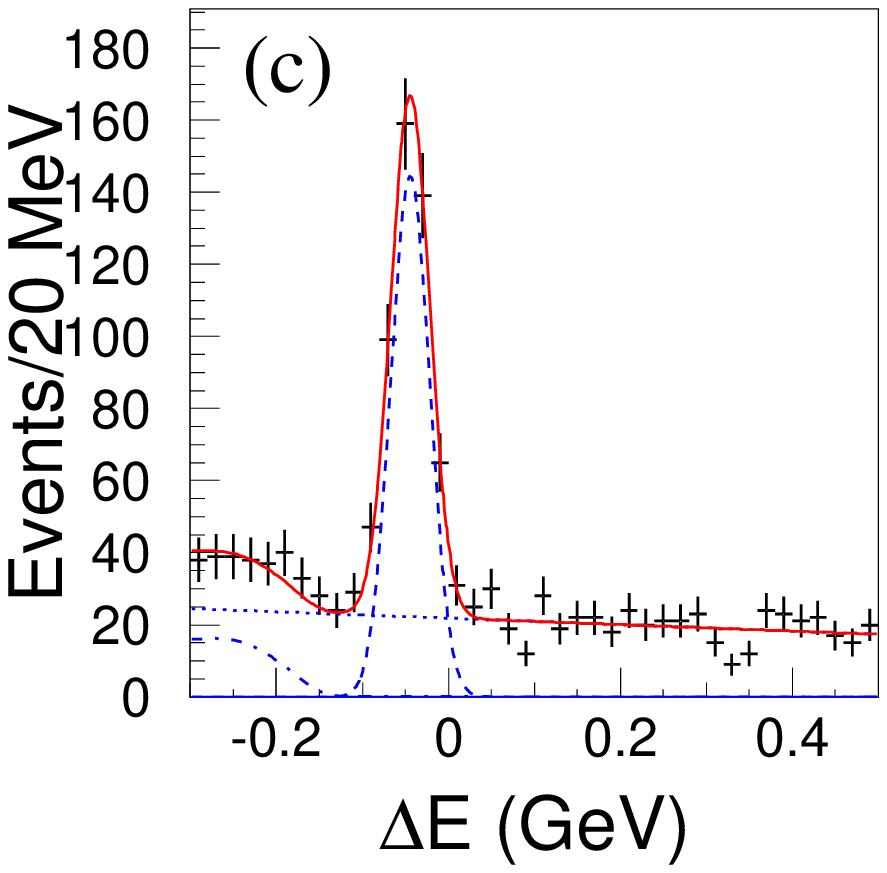}}
\resizebox{0.4\textwidth}{!}{\includegraphics{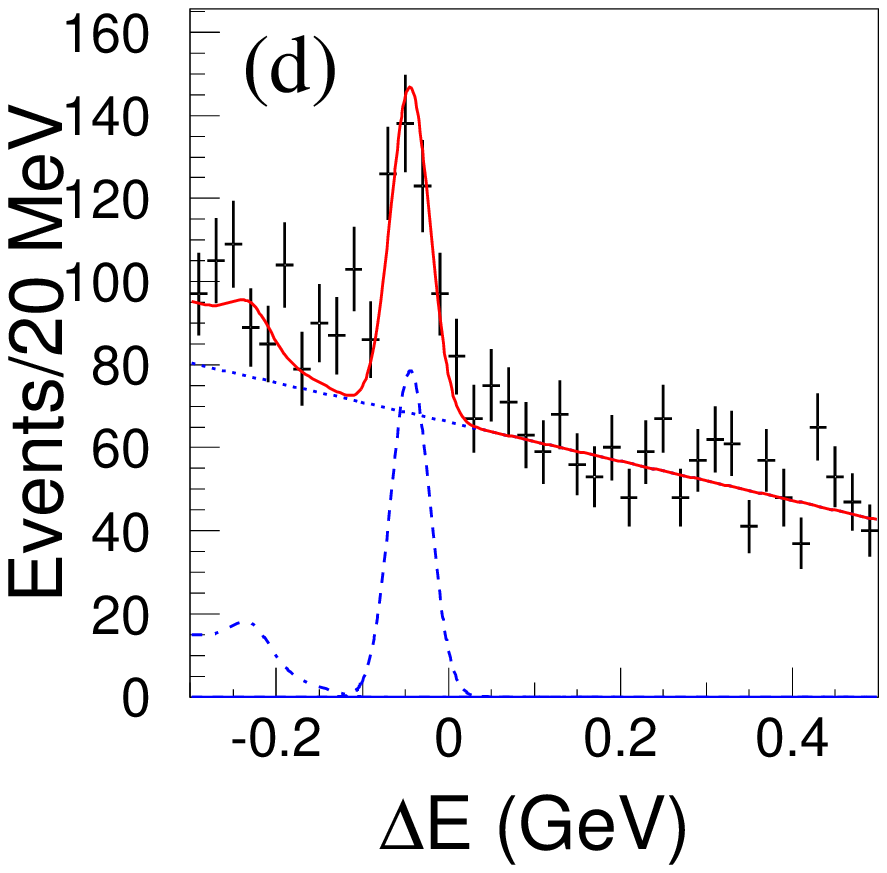}}
\end{center}
\caption{$\Delta E$ distributions in the $M_{\rm bc}$ signal region for 
(a) $B^0 \rightarrow \pi^+\pi^-$ candidates with $LR$ $>$ 0.825,
(b) $B^0 \rightarrow \pi^+\pi^-$ candidates with $LR$ $\leq$ 0.825,
(c) $B^0 \rightarrow K^+\pi^-$ candidates with $LR$ $>$ 0.825, and
(d) $B^0 \rightarrow K^+\pi^-$ candidates with $LR$ $\leq$ 0.825.
The sum of the signal and background functions is shown as a solid curve.
The solid curve with hatched area represents the $\pi^+\pi^-$ component,
the dashed curve represents the $K^+\pi^-$ component,
the dotted curve represents the continuum background, and
the dot-dashed curve represents the charmless three-body $B$ decay
background component.}
\label{fig:DeltaE}
\end{figure}
%%%%%%%%%%%%%%%%%%%%%%%%
%
\section{SIGNAL YIELD}
\label{sec:yield}
Figures~\ref{fig:DeltaE}(a) and (b) show the $\Delta E$ distributions 
for the $B^0$ $\rightarrow$ $\pi^+\pi^-$ candidates 
that are in the $M_{\rm bc}$ signal region 
with $LR$ $>$ 0.825 and with $LR$ $\leq$ 0.825, respectively, 
after flavor tagging and vertex reconstruction.
In the $M_{\rm bc}$ and $\Delta E$ signal region, 
we find 275 candidates for $LR$ $>$ 0.825 and
485 candidates for $LR$ $\leq$ 0.825.
The $B^0$ $\rightarrow$ $\pi^+\pi^-$ signal yield for $LR$ $>$ 0.825
is extracted by fitting the $\Delta E$ distribution
with a Gaussian signal function
plus contributions from misidentified $B^0 \to K^+\pi^-$ events, 
three-body $B$-decays, and continuum background. The fit yields 
$106^{+16}_{-15}$ $\pi^+\pi^-$ events, 
$41^{+10}_{-9}$ $K^+\pi^-$ events and 
$128^{+5}_{-6}$ continuum events in the signal region.
The errors do not include systematic uncertainties unless otherwise stated.
%%v4<<
%%v5<< 
%%v5Here the error of the yield of continuum events does not include 
%%v5Poisson fluctuation.
Here the error on the yield of continuum events in the signal region
is obtained by scaling the error of the yield from the fit that encompasses
the entire $\Delta E$ range.
%%v5>>
%%v4 For the fit to the events with $LR\leq 0.825$,
%%v4 we fix the level of $K^+\pi^-$ background 
%%v4 by scaling the $LR>0.825$ number by a MC-determined factor
%%v4 and that of the continuum background from the sideband data.
For $LR\leq 0.825$, 
we fix the level of $\pi^+\pi^-$ signal by scaling the $LR>0.825$ number
by a MC-determined factor and 
that of the continuum background from the sideband.
The ratio of the $K^+\pi^-$ background to the $\pi^+\pi^-$ signal is fixed to 
the value measured with the $LR> 0.825$ sample.
%%v4>>
We obtain $57\pm 8$ $\pi^+\pi^-$ events, 
$22^{+6}_{-5}$ $K^+\pi^-$ events and 
$406\pm 17$ continuum events in the signal region for $LR \leq 0.825$.
The contribution from
three-body $B$-decays is negligibly small in the signal region.
Figures~\ref{fig:DeltaE}(c) and (d) show the $\Delta E$ distributions for
the selected $B^0 \rightarrow K^+\pi^-$ candidates.

\section{MAXIMUM LIKELIHOOD FIT}
\label{sec:MLH}

The proper-time interval resolution function $R_{\pi\pi}$ for
$B^0$ $\rightarrow$ $\pi^+\pi^-$ signal events is formed by convolving four components:
the detector resolutions for $z_{CP}$ and $z_{\rm tag}$,
the shift in the $z_{\rm tag}$ vertex position due to secondary tracks originating
from charmed particle decays,
and the smearing due to the kinematic approximation used to convert $\Delta z$
to $\Delta t$.
We use the same parameters as those used for 
the sin2$\phi_1$ measurement~\cite{CP1_Belle3}.
We determine resolution parameters from fitting the data for the neutral and 
charged $B$ meson lifetimes.
A small component of broad outliers in the $\Delta z$ distribution,
caused by mis-reconstruction, is represented by a Gaussian function.
The width of the outlier component is determined to be $42^{+5}_{-4}$~ps;
the fractions $f_{ol}$ of the outlier components are $(2\pm1)\times10^{-4}$ 
for events with both vertices reconstructed with more than one track, and
$(2.7\pm0.2)\times10^{-2}$ for events with one or two single-track vertices.
We assume $R_{\pi\pi}$ = $R_{K\pi}$ and denote them
collectively as $R_{sig}$.
The parameters of the continuum background resolution function 
$R_{q\overline{q}}$ are obtained from the sideband data.

The $CP$ asymmetry parameters, $\apipif$ and $\spipif$,
are obtained from an unbinned maximum likelihood fit
to the observed proper-time distribution. 
For this purpose, we 
use probability density functons (PDFs) that are
based on theoretical distributions that are
diluted and smeared by the detector response.
The PDF for $B^0$ $\rightarrow$ $\pi^+\pi^-$ signal events (${\cal P}^q_{\pi\pi}$) 
is given by
Eq.(\ref{eq:R_q}), with $q$ replaced by $q(1-2w_l)$ to account for the dilution
due to wrong flavor tagging.
The PDF for 
$B^0$ $\rightarrow$ $K^+\pi^-$ background events is
${\cal P}^q_{K\pi}(\Delta t,w_l)
  =
  { e^{-|\Delta t|/\tau_{B^0}} }/{(4\tau_{B^0})}
 \{
 1 + q\cdot (1-2w_l)\cdot{\akpif}
 \cdot\cos(\Delta m_d\Delta t)
 \},$
where we assume as a default that there is no $CP$ asymmetry for the $B^0$ $\rightarrow$ $K^+\pi^-$ mode. 
The effect of a possible non-zero value for $\akpif$ 
is determined by varying $\akpif$
by the error obtained from fits to the self-tagged 
$B^0$ $\rightarrow$ $K^+\pi^-$ sample and is included in
the systematic error.
The PDF for continuum background events is 
${\cal P}_{q\overline{q}}(\Delta t)
 =
(1+q{\cdot}{\abkgf})/{2}\{f_\tau { e^{-|\Delta t|/\tau_{\rm bkg}} }/{(2\tau_{\rm bkg})}
 + ( 1 - f_\tau )\delta(\Delta t)\},$
where $f_\tau$ is the fraction of the background with effective lifetime
$\tau_{\rm bkg}$ and $\delta$ is the Dirac delta function. 
%%v4<< For $B^0$ $\rightarrow$ $\pi^+\pi^-$ events where both vertices
For $B^0$ $\rightarrow$ $\pi^+\pi^-$ candidates where both vertices
%%v4>>
have at least two tracks,
we use $f_\tau$ = 0.014 $^{+0.006}_{-0.004}$ and 
$\tau_{\rm bkg}=2.37^{+0.44}_{-0.34}~{\rm ps}$, which are
determined from the events in the
$q{\overline{q}}$-background-dominated 
%%v4<<
%%v4 $\Delta E$ vs. $\MBC$ sideband region. 
sideband region: $5.20~{\rm GeV/c^2} < \MBC < 5.26~{\rm GeV/c^2}$ and
$0.10~{\rm GeV} < \Delta E < 0.50~{\rm GeV}$. 
%%v4>>
For events with
a single-track vertex,
we use $f_{\tau}$ = 0.
The effect of the uncertainty in $\abkgf$,
determined by varying $\abkgf$ by
the error from the fit to the sideband data,
is included in the systematic error. 

We define the likelihood value for each ($i$-th) event as 
a function of $\apipif$ and $\spipif$:
\begin{eqnarray}
{P_i =
(1 - f_{ol})\int^{+\infty}_{-\infty}
\{
(f^m_{\pi\pi}{\cal P}^q_{\pi\pi}(\Delta t^\prime, w_l; \apipif, \spipif)} \nonumber \\
+ f^m_{K\pi}{\cal P}^q_{K\pi}(\Delta t^\prime, w_l))
\cdot R_{sig}(\Delta t_i-\Delta t^\prime) \nonumber \\
  + f^m_{q\overline{q}}{\cal P}_{q\overline{q}}(\Delta t^\prime)
\cdot R_{q\overline{q}}(\Delta t_i-\Delta t^\prime)\}
d\Delta t^\prime 
+ f_{ol}{\cal P}_{ol}(\Delta{t_i}).
\label{eq:likelihood}
\end{eqnarray}
Here the probability functions $f^m_k$ ($k$ = $\pi\pi$, $K\pi$ or 
$q\overline{q}$ )
are determined on an event-by-event basis as functions of  $\Delta E$ 
and $\MBC$ for each $LR$-$r$ interval ($m$= 1, 12). 
For example, $f^m_{\pi\pi}(\Delta E, \MBC)$ is 
${F_{\pi\pi}{g^m_{\pi\pi}}}
  /({F^m_{q\overline{q}}{g^m_{q\overline{q}}}+
   F_{\pi\pi}{g^m_{\pi\pi}}+F_{K\pi}{g^m_{K\pi}}})$,
where $g^m_k$ is the average fraction of event-type $k$ for the $m$-th
$LR$-$r$ interval 
($g^m_{\pi\pi}$ + $g^m_{K\pi}$ + $g^m_{q\overline{q}}$ = 1).
We determine these parameters from the numbers of events in the sideband
data and from fractions of $\bz \to \pi^+\pi^-$ MC events.
Table~\ref{tbl:frac} lists 
the values of $g^m_k$ for the 12 $LR$-$r$ regions.
We obtain $g^m_{K\pi}=0.382\times g^m_{\pi\pi}$
 from the fit to the $\Delta E$ distribution
for the $B^0 \rightarrow \pi^+\pi^-$ candidates with $LR > 0.85$.
The distributions of $\Delta E$ and $\MBC$ for the $B^0$ $\rightarrow$ $\pi^+\pi^-$
 signal shape function $F_{\pi\pi}(\Delta E, \MBC)$ and 
$B^0$ $\rightarrow$ $K^+\pi^-$ background shape function 
$F_{K\pi}(\Delta E, \MBC)$ are fit with Gaussian functions. 
$F^m_{q\overline{q}}(\Delta E, \MBC)$ is
the continuum background shape function, and the distributions of
$\Delta E$ and
$\MBC$ are fit with $m$-dependent linear functions and the ARGUS background 
function~\cite{ARGUS}, respectively.
The small number of signal and background 
events that have large $\Delta{t}$ 
are accommodated by the
outlier PDF, ${\cal P}_{ol}$, with
fractional area $f_{ol}$.

In the fit,
$\spipif$ and $\apipif$ are free parameters 
determined by maximizing
the likelihood function
${\cal L}=\prod_i P_i$, where the product is over all
$B^0 \rightarrow \pi^+\pi^-$ candidates.

%%%%%%%%%%%%%%%%%%%%%%%%%%%%%%%%%%%%%%%%%%%%%%%%%%%%%%%%%%%%%%%%%%%%%%%%%%%
\begin{table}[!htb]
\caption{The fractions of expected
$B^0$ $\rightarrow$ $\pi^+\pi^-$ and continuum events for
the 12 $LR$-$r$ regions. 
}
\label{tbl:frac}
\begin{ruledtabular}
\begin{tabular}{ccc|cc}
$m$ & $r$ interval & $LR$ interval &  $g^m_{\pi\pi}$  & $g^m_{q\overline{q}}$ \\
 \hline
1 & $0.000 - 0.250$  & $0.825 - 1.000$ & $0.296\pm0.077$ & $0.591\pm0.028$ \\
2 & $0.250 - 0.500$  & $0.825 - 1.000$ & $0.385\pm0.094$ & $0.468\pm0.026$ \\
3 & $0.500 - 0.625$  & $0.825 - 1.000$ & $0.407\pm0.134$ & $0.438\pm0.027$ \\
4 & $0.625 - 0.750$  & $0.825 - 1.000$ & $0.442\pm0.110$ & $0.389\pm0.024$ \\
5 & $0.750 - 0.875$  & $0.825 - 1.000$ & $0.522\pm0.081$ & $0.279\pm0.022$ \\
6 & $0.875 - 1.000$  & $0.825 - 1.000$ & $0.670\pm0.129$ & $0.074\pm0.009$ \\
7 & $0.000 - 0.250$  & $0.525 - 0.825$ & $0.087\pm0.034$ & $0.880\pm0.040$ \\
8 & $0.250 - 0.500$  & $0.525 - 0.825$ & $0.127\pm0.049$ & $0.824\pm0.040$ \\
9 & $0.500 - 0.625$  & $0.425 - 0.825$ & $0.124\pm0.036$ & $0.829\pm0.041$ \\
10 & $0.625 - 0.750$  & $0.425 - 0.825$ &$0.129\pm0.050$ & $0.822\pm0.040$ \\
11 & $0.750 - 0.875$  & $0.425 - 0.825$ &$0.170\pm0.060$ & $0.765\pm0.040$ \\
12 & $0.875 - 1.000$  & $0.325 - 0.825$ &$0.390\pm0.098$ & $0.461\pm0.032$ \\
\end{tabular}
\end{ruledtabular}
\end{table}
%%%%%%%%%%%%%%%%%%%%%%%%%%%%%%%%%%%%%%%%%%%%%%%%%%%%%%%%%%%%%%%%%%%%%%%%%%%

We check the validity of our fitting procedure 
with a large ensemble of MC pseudo-experiments wherein events are generated 
with nominal PDFs and the observed number of events.
The parameters in the PDFs are taken from data.
For various input values of
$\spipi$ and $\apipi$, we confirm that there is no bias in the fit.
The MC pseudo-experiments are described in detail in Appendix~\ref{app:ToyMC}.

\section{FIT RESULTS}
\label{sec:result}
The unbinned maximum likelihood fit to the
760 $B^0$ $\rightarrow$ $\pi^+\pi^-$ candidates 
(391 $B^0$- and 369 $\bzb$-tags), 
containing 163$^{+24}_{-23}$ $\pi^+\pi^-$ signal events, yields
$\apipif$ = $\avalue$ and $\spipif$ = $\svalue$.
In Figs.~\ref{fig:asym}(a) and (b), we show the raw, unweighted
$\Delta t$ distributions for
the 148 $B^0$- and 127 $\bzb$-tagged events with $LR$ $>$ 0.825.
The fit curves use $\apipif$ and $\spipif$ values that are obtained 
from all of the $LR$-$r$ regions.
The background-subtracted $\Delta t$ distributions
are shown in Fig.~\ref{fig:asym}(c).
Figure~\ref{fig:asym}(d) shows the background-subtracted
$CP$ asymmetry between
the $B^0$- and
$\bzb$-tagged events 
as a function of $\Delta t$. The result
of the fit is superimposed and is shown by the solid curve.
%%%%%%%%%%%%%%%%%%%%%%
\begin{figure}[!htbp]
\begin{center}
\resizebox{0.6\textwidth}{!}{\includegraphics{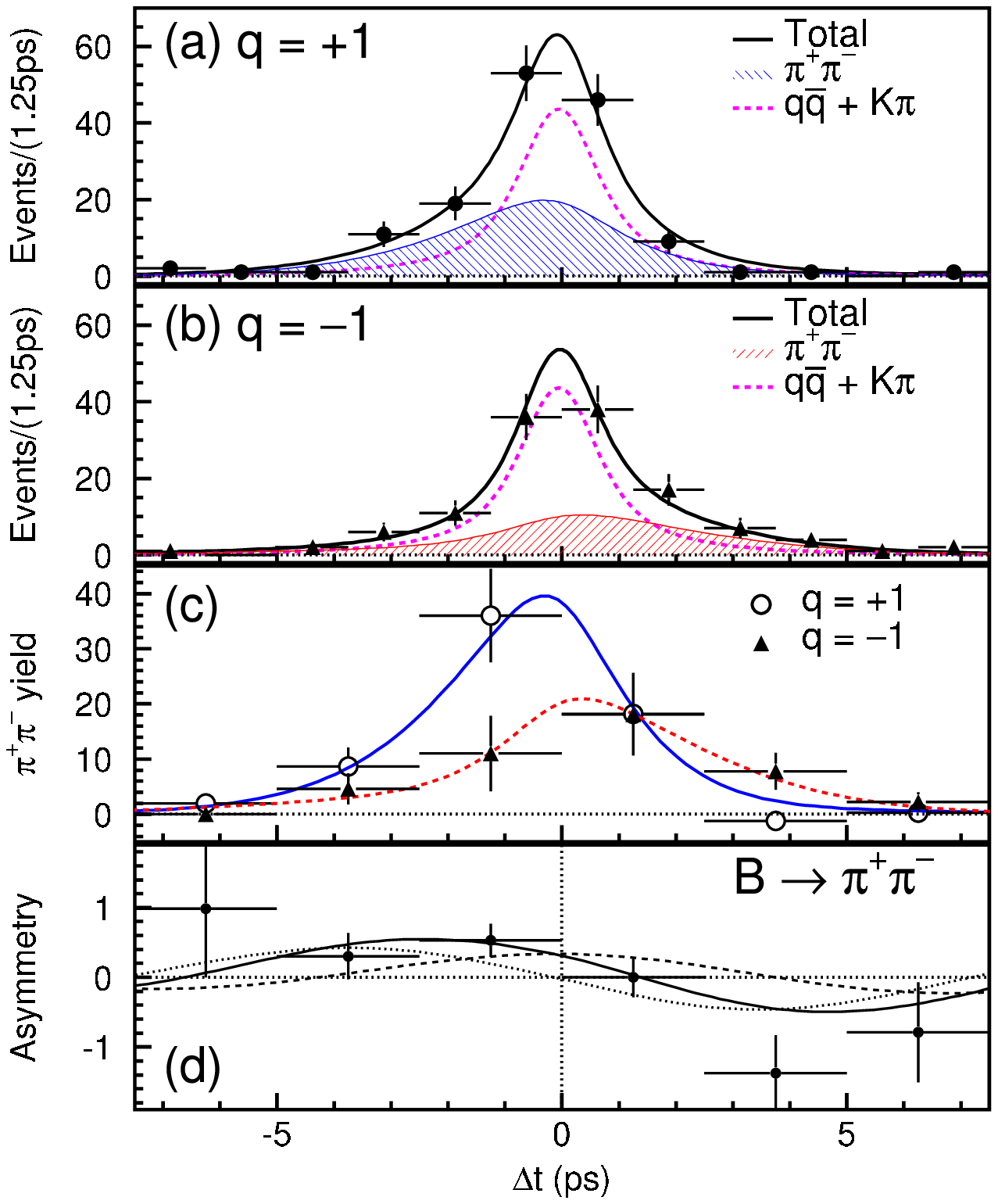}}
\end{center}
\caption{
The raw, unweighted $\Delta t$ distributions for 
the 275 $B^0 \rightarrow \pi^+\pi^-$ candidates with $LR$ $>$ 0.825 in the
signal region: 
(a) 148 candidates with $q = +1$, i.e. the tag side is identified as $B^0$;
(b) 127 candidates with $q = -1$; 
(c) $B^0$ $\rightarrow$ $\pi^+ \pi^-$ yields after background subtraction. The errors
are statistical only and do not include the error on
the background subtraction;
(d) the $CP$ asymmetry for $B^0 \rightarrow \pi^+\pi^-$
after background subtraction.
In Figs. (a) through (c), the curves show the
results of the unbinned maximum likelihood fit to the $\Delta t$
distributions of the 760 $B^0$ $\rightarrow$ $\pi^+\pi^-$ candidates.
In Fig. (d), the solid curve shows the resultant $CP$ asymmetry,
while the dashed (dotted) curve is the contribution from
the cosine (sine) term.
}
\label{fig:asym}
\end{figure}
%%%%%%%%%%%%%%%%%%%%%%%%%

We test the goodness-of-fit from a $\chi^2$
comparison of the results of the unbinned fit and
the $\Delta t$ projections for $B^0 \rightarrow \pi^+\pi^-$ candidates~\cite{footnoteChi2}.
We obtain $\chi^2$ = 
10.9/12~DOF (13.3/12~DOF)
for the $\Delta t$ distribution of
the $B^0$ ($\overline{B}^0$) tags.

As shown in Table~\ref{tbl:fraction},
an ensemble of MC pseudo-experiments indicates a 16.6$\%$ probability 
to measure $CP$ violation at or above the one we observe when the input values
are $\apipi = \avaluecnst$ and $\spipi = \svaluecnst$, 
which correspond to the values at the point of maximum likelihood 
in the physically allowed region ($\spipi^2+\apipi^2\leq 1$); in this measurement it is
located at the physical boundary ($\apipi^2 + \spipi^2 = 1$).
This set of MC pseudo-experiments also indicates that 
for an input value on the physical boundary,
the probability of obtaining a result outside the physically allowed region
is large (60.1\%).
%%%%%%%%%%%%%%%%%%%%%%%%%%%%%%%%%%%%%%%%%%%%%%%%%
\begin{table}[!htbp]
\caption{The fractions of MC pseudo-experiments outside the physical boundary and
above the $CP$ violation we observe for various input values.
${\rho}_{\pi\pi} = \sqrt{\apipi^2+\spipi^2}$.
The selected points are on the line segment between ($\apipi, \spipi$) = 
(0,0) and ($+0.57, -0.82$).
}
\label{tbl:fraction}
\begin{ruledtabular}
\begin{tabular}{ccc}
  Input ${\rho}_{\pi\pi}$ & The fractions outside & The fractions above\\ 
   & the physical boundary &  the $CP$ violation  \\ 
   &  ($\%$)  & we observe ($\%$) \\ \hline
0.00 &  1.8 &  0.07 \\
0.20 &  3.3 &  0.17 \\
0.40 &  7.3 &  0.62 \\
0.60 & 16.4 &  1.7 \\
0.80 & 34.4 &  6.0 \\ 
1.00 & 60.1 & 16.6 \\ 
\end{tabular}
\end{ruledtabular}
\end{table}
%%%%%%%%%%%%%%%%%%%%%%%%%%%%%%%%%%%%%%%%%%%%%%%%%
%
\subsection{Statistical errors}
\label{sec:stat_error}

As described below in Section~\ref{sec:CL},
we obtain confidence intervals for $\apipi$ and $\spipi$
with a frequentist approach
where we use MC pseudo-experiments to determine acceptance regions,
and we quote the rms values of the MC $\apipif$ and $\spipif$
%%v5 distributions as the errors of our measurement.
distributions as the statistical errors of our measurement.
We obtain 
$\apipif=\avalue\atoyerr\text{(stat)}$ and
$\spipif=\svalue\stoyerr\text{(stat)}$.
Here we choose values at the point of maximum likelihood in the physically allowed region,
($\apipi$, $\spipi$) = ($\avaluecnst$, $\svaluecnst$),
for the input to the MC pseudo-experiments used
to obtain the statistical errors.
The rms values determined with input values of 
($\apipi,\spipi)$ = $(0,0)$ are slightly different;
for these input values we obtain $\pm 0.28$ and $\pm 0.39$ 
for the $\apipif$ and $\spipif$ errors, respectively.

In the literature, the statistical error is 
usually determined from the parameter dependence
of the log-likelihood ratio $-$2ln(${\cal L}/{\cal L}_{\rm max})$
that is obtained from the fit. Here we call this estimator the MINOS error, 
which corresponds to the deviation from
the best fit parameter when $-2{\rm ln}({\cal L}/{\cal L}_{\rm max})$ is
changed by one.
The MINOS error is a convenient approximation for
defining a 68.3\% (1$\sigma$) confidence interval;
however, care is needed when defining intervals
at higher confidence levels. 
Figure~\ref{fig:log-LH} shows the log-likelihood ratio curves from our
data, where deviations from parabolic behavior are evident;
for example, 
%%v5<<
%%v5 3$\sigma$ from the MINOS error for $\spipif$ is
3$\sigma$ from the MINOS error for the positive side of $\spipif$
%%v7 , which is relevant to establishing the significance of the non-zero $\spip%%v7 i$ value, 
%%v5>>
is considerably smaller than a three-standard-deviation error 
defined by the deviation from the best fit
parameter when $-$2ln(${\cal L}/{\cal L}_{\rm max}$) is changed by 9.
%%v7<<
Thus, the MINOS error overestimates the significance of a non-zero $\spipi$
value.
%%v7>>
The MINOS errors obtained from the curves are also smaller
than the expectations from the MC pseudo-experiments, as shown 
in Fig.~\ref{fig:extream_error};
the probability of obtaining a MINOS error smaller than that in our measurement is
1.2\% (12.0\%) for $\spipif$ ($\apipif$)~\cite{footnoteStatError}.
These characteristics are reproduced in a fraction of
the MC pseudo-experiments that have
$\apipi$ and $\spipi$ input values that are  
close to the physical boundary.
%%v4<<
%%v4 We therefore conclude that the errors derived from the
We therefore conclude that the rms values of the distributions
of fit outputs of $\apipi$ and $\spipi$ from the
%%v4>>
MC pseudo-experiments, rather than the MINOS errors, are
%%v5 more appropriate as the standard errors for this measurement.
more appropriate as the standard statistical errors for this measurement.
We describe an investigation of the source of 
the small MINOS errors in Appendix~\ref{app:small-error}.

%%%%%%%%%%%%%%%%%%%%%%%%%%%%%%%%%%%%%%%%%%%%%%%%%%%
\begin{figure}[!htb]
\begin{center}
\resizebox{0.4\textwidth}{!}{\includegraphics{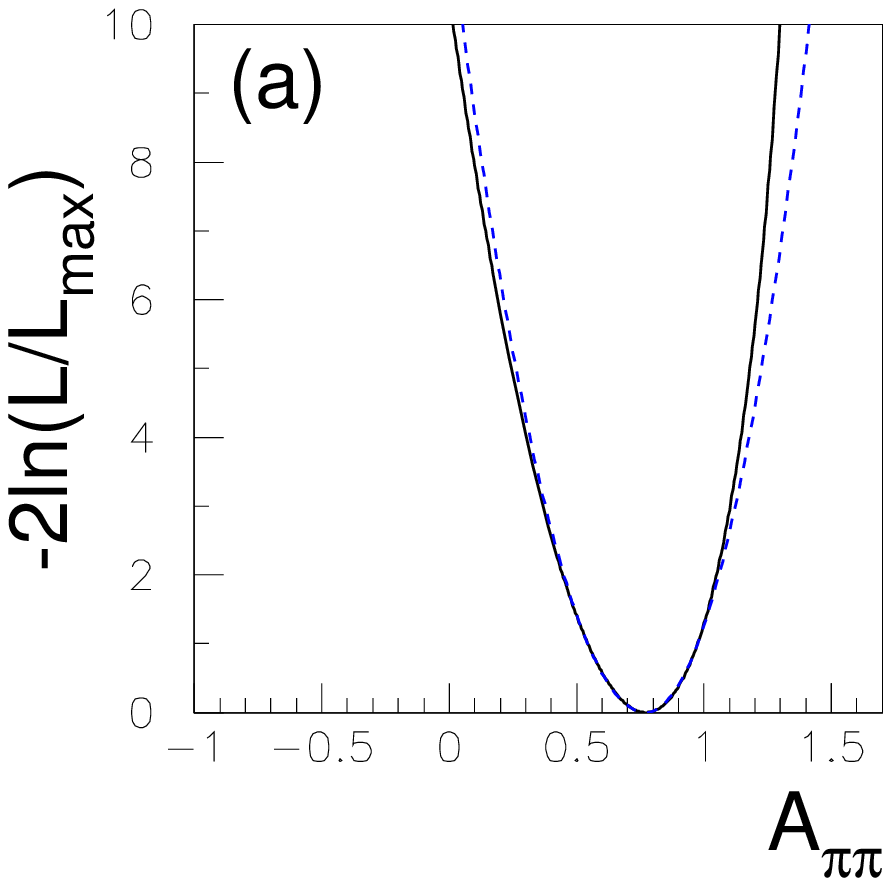}}
\resizebox{0.4\textwidth}{!}{\includegraphics{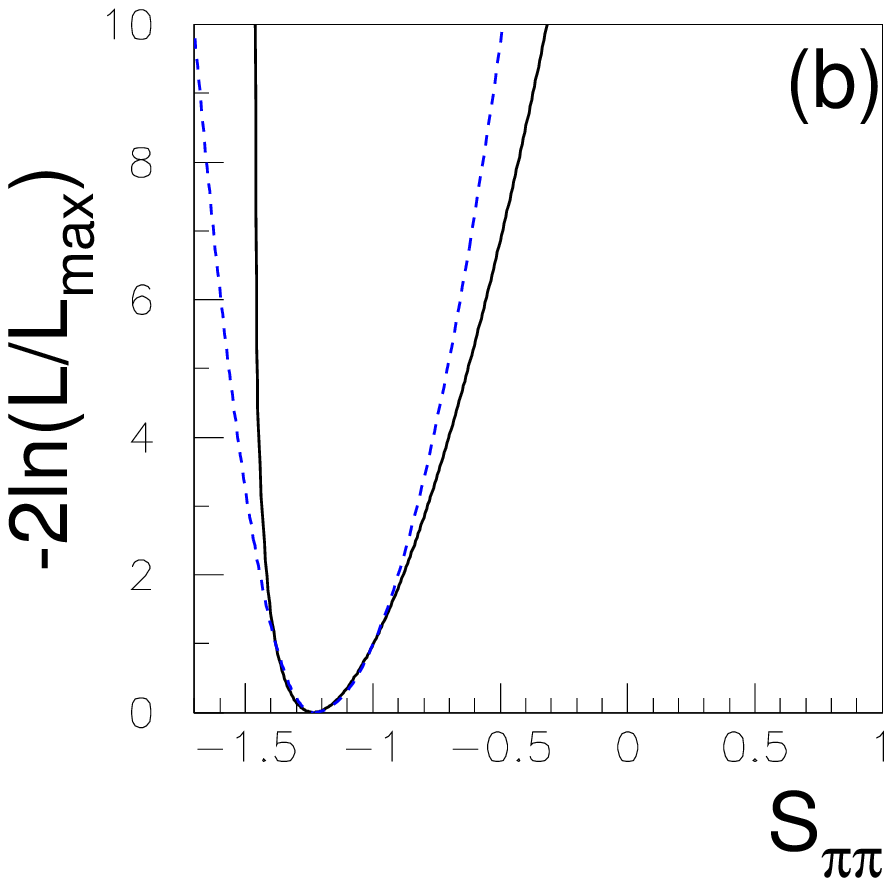}}
\caption{(a) The value of $-$2ln(${\cal L}/{\cal L}_{\rm max}$) vs. $\apipif$ and 
(b) the value of 
$-$2ln(${\cal L}/{\cal L}_{\rm max}$) vs. $\spipif$.
The dotted curves represent parabolic functions which pass the point at 1$\sigma$.
}
\label{fig:log-LH}
\end{center}
\end{figure}
%%%%%%%%%%%%%%%%%%%%%%%%%%%%%%%%%%%%%%%%%%%%%%%%%%%%

%%%%%%%%%%%%%%%%%%%%%%%%%%%%%%%%%%%%%%%%%%%%%%%%%%%%%%%%%%%%%%%%%%%%%%%%%%%
\begin{figure}[!htbp]
\begin{center}
\resizebox{0.6\textwidth}{!}{\includegraphics{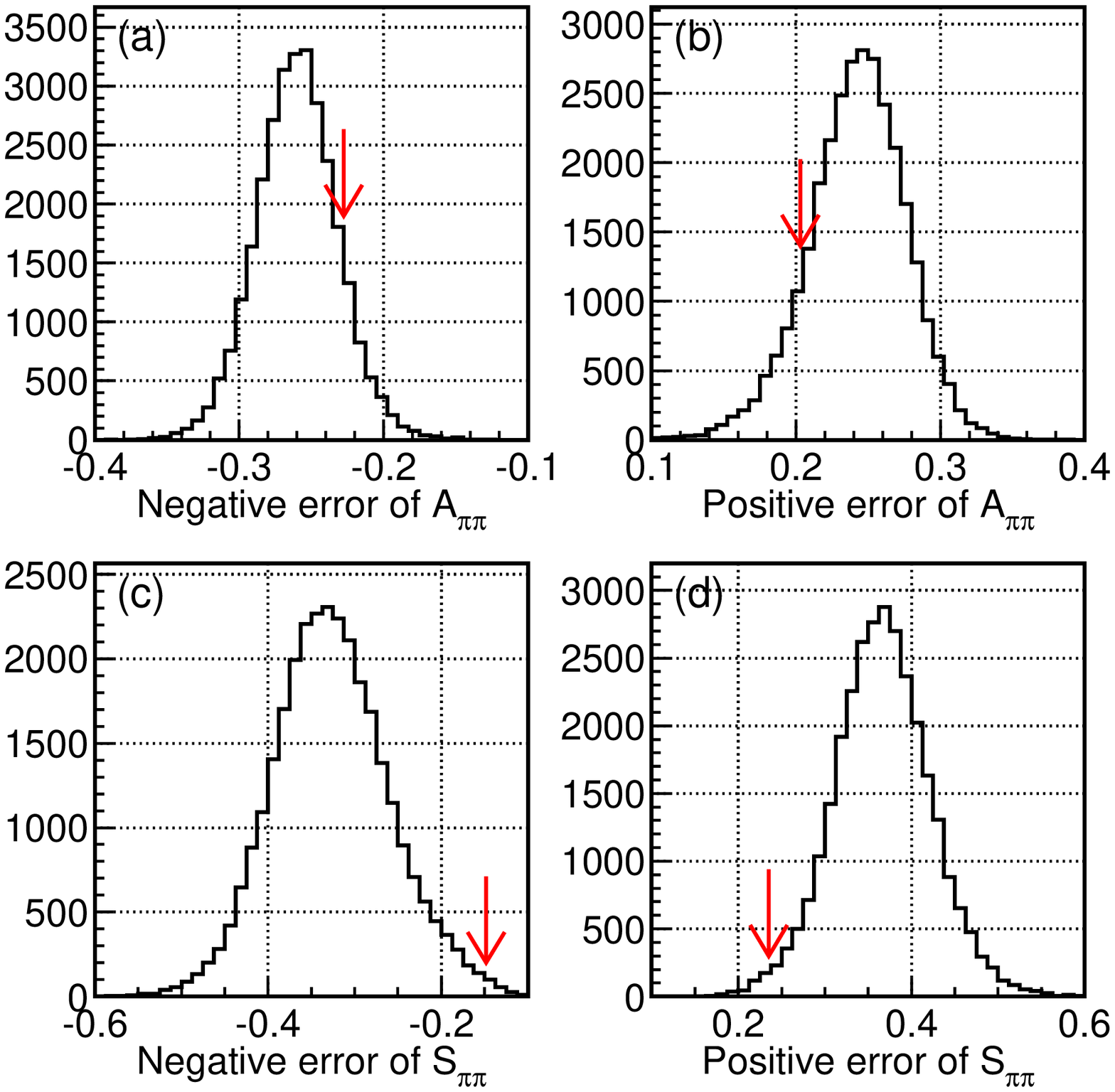}}
\caption{The result of MC pseudo-experiments with input values of
$\apipi = \avaluecnst$ and $\spipi = \svaluecnst$: 
the distributions of (a) the negative and (b) positive MINOS errors of $\apipif$, and
(c) the negative and (d) positive MINOS errors of $\spipif$.
The arrows indicate the MINOS errors obtained from the fit to data.
} 
\label{fig:extream_error}
\end{center}
\end{figure}
%%%%%%%%%%%%%%%%%%%%%%%%%%%%%%%%%%%%%%%%%%%%%%%%%%%%%%%%%%%%%%%%%%%%%%%%%

\subsection{Systematic errors}
\label{sec:syst}
The sources of the systematic error are listed in Table~\ref{tbl:syst}.
We add each contribution in quadrature for the
total systematic errors. We obtain
\begin{eqnarray*}
\apipif &=& \aresult, \\
\spipif &=& \sresult.
\end{eqnarray*}
The systematic error on $\apipif$ is primarily due to
uncertainties in the background fractions and the vertexing. 
For $\spipif$, the background fractions and
a possible fit bias near the physical boundary 
are the two leading components.
Below we explain each item in order.
%%%%%%%%%%%%%%%%%%%%%%%%%%%%%%%%%%%%%%%%%%%%%%%%%%%%%%%%%%%%%%%%%%%%%%%%
\begin{table}[!hbtp]
 \begin{center}
 \caption{Systematic errors for $\apipif$ and $\spipif$.}
 \label{tbl:syst}
\begin{ruledtabular}
  \begin{tabular}{l|ll|ll}
  & \multicolumn{2}{c|}{$\apipif$}  &  \multicolumn{2}{c}{$\spipif$} \\ 
   Source                   &  $+$error  & $-$error &  $+$error  & $-$error \\   \hline
 Background fractions           & +0.058 & $-$0.048 & +0.044 & $-$0.055 \\
 Vertexing                      & +0.044 & $-$0.054 & +0.037 & $-$0.012 \\
 Fit bias                       & +0.016 & $-$0.021 & +0.052 & $-$0.020 \\
 Wrong tag fraction             & +0.026 & $-$0.021 & +0.015 & $-$0.016 \\
 Physics ($\tau_{B^0}$, $\Delta m_d$, 
             ${\cal A}_{K\pi}$) & +0.021 & $-$0.014 & +0.022 & $-$0.022 \\
 Resolution function            & +0.019 & $-$0.020 & +0.010 & $-$0.013 \\
 Background shape               & +0.003 & $-$0.015 & +0.007 & $-$0.002 \\
   \hline                                                  
  Total                         & +0.084 & $-$0.083 & +0.083 & $-$0.067\\
\hline
  \end{tabular}
\end{ruledtabular}
 \end{center}
\end{table}
\subsubsection{Background fractions}
We estimate the systematic errors that arise
from uncertainties in the parameters
used for the event-by-event
background fractions $f^m_{K\pi}$ and $f^m_{q\overline{q}}$
as well as the signal fraction $f^m_{\pi\pi}$.
Parameters that are determined from data
are varied by their errors and fits are repeated;
we add the contribution from each variation in quadrature.

As explained in Sec.~\ref{sec:MLH},
we rely on a MC $\bz \to \pi^+\pi^-$ sample
to determine $g^m_k$, the background fraction
in each $LR$-$r$ region $m$.
We measure the regional event fractions 
in $\bz \to D^{(*)}\pi$ control samples,
and compare the results with those
in the MC $\bz \to \pi^+\pi^-$ sample.
Each $g^m_k$ value is then modified by an amount determined
from the difference between data and MC,
and from the statistical error in the 
control samples. We repeat the fit to obtain $\apipi$ and $\spipi$,
and add each difference from the nominal value in quadrature.

The $K\pi$ background yield is obtained from the fit
to the $\Delta E$ distribution.
We estimate the systematic error associated with this method
from an independent yield measurement
based on a $K\pi$ enriched-sample and the $K/\pi$ separation performance,
which will be described in Sec.~\ref{sec:crosscheck}.

The PDF for continuum background used in the fit
assumes no asymmetry (${\cal A}_{\rm bkg} = 0$)
between the number of events with $q=+1$
and with $q=-1$.
We estimate the systematic error due to this assumption
by varing ${\cal A}_{\rm bkg}$ by $\pm 0.02$, 
based on the measurement ${\cal A}_{\rm bkg} = 0.013\pm 0.006$
from the sideband data.

\subsubsection{Vertexing}
We search for possible biases that may arise from
the track and vertex selection by 
repeating the analysis with modified selection criteria.
We include the observed changes in the systematic error.
We also repeat the analysis by introducing charge-dependent
shifts in the $z$ direction 
artificially, and include 
the resulting change in the systematic error.
Here the amount of the shift is determined
from studies with cosmic rays and with 
the two-photon process $e^+e^- \to \pi^+\pi^-\pi^+\pi^-$.
The systematic error associated with the IP profile
is estimated by varying 
the IP smearing that is used to account for the $B$ flight length.

\subsubsection{Fit bias and other sources}
We use large-statistics MC pseudo-experiments to determine
the systematic error due to possible fit biases 
for the input $\apipi$ and $\spipi$ values near the physical boundary. 
We also perform a fit to MC $\bz \to \pi^+\pi^-$ events 
that are generated by using 
%HS186 a Geant-based simulation. We obtain results that are consistent with input
a GEANT-based simulation. We obtain results that are consistent with input
values within the statistical errors, which are conservatively
included in the systematic error.

Systematic errors due to uncertainties in the wrong tag fractions
are estimated by varying each wrong tag fraction in
each $r$ region, and repeating the fit procedure.
We also repeat the fit using wrong tag fractions
obtained for $\bz$- and $\bzb$-tagged control samples separately.
We add each contribution in quadrature. 

We estimate the systematic errors associated with
parameters in the resolution functions, in the background PDF, and
the physics parameters ($\tau_{B^0}$, $\Delta m_d$, and ${\cal A}_{K\pi}$)
by repeating the fit varying these parameters by their errors.

\subsection{Crosschecks}
\label{sec:crosscheck}
We perform a number of crosschecks. 
We measure the $B$ meson lifetime using the same vertex reconstruction method.
The results of the application of the
same analysis to various subsamples are also examined.
In addition, we check for biases in the analysis
using samples of non-$CP$ eigenstates, $B^0 \rightarrow
K^+\pi^-$ decays, and sideband data. 

We perform a $B^0$ lifetime measurement with
the $B^0 \rightarrow \pi^+\pi^-$ candidate events that 
uses the same background fractions, vertex reconstruction methods,
and resolution functions that are used for the $CP$ fit. 
Figure~\ref{fig:lifetime}(a) shows the fit result.
The fit to the events in the $\pi^+\pi^-$ sideband
is also shown in Fig.~\ref{fig:lifetime}(c); the fit curve
is used for the PDF of the continuum background.
%%%%%%%%%%%%%%%%%%%%%%%%
\begin{figure}[!htbp]
\begin{center}
\resizebox{0.6\textwidth}{!}{\includegraphics{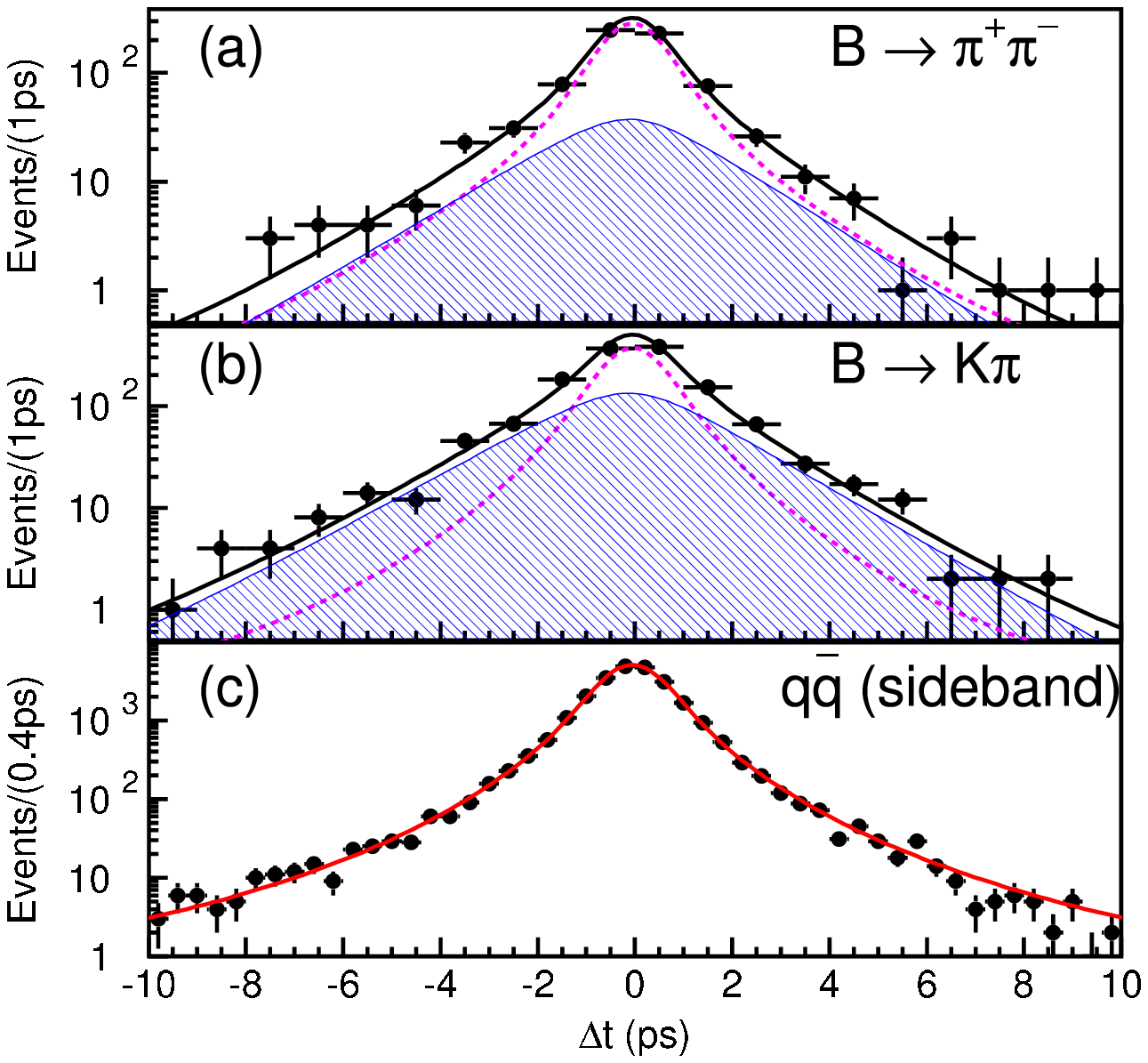}}
\end{center}
\caption{Results of the lifetime fits for (a) $\pi^+\pi^-$ candidates
and (b) $K^+\pi^-$ candidates. 
The solid curves are the results of the fits, the shaded areas
are the signal and the dashed curves are background contributions.
(c) The fit to events in the $\pi^+\pi^-$ sideband.}
\label{fig:lifetime}
\end{figure}
%%%%%%%%%%%%%%%%%%%%%%%%
The result, 
$\tau_{B^0} = 1.42^{+0.14}_{-0.12}{\rm ~ps}$,
is consistent with the world-average value~\cite{PDG}.

We repeat the fits for 
$\apipi$  and $\spipi$ with
$\pi^+\pi^-$ candidate samples selected
with more stringent selection criteria.
The $K^+\pi^-$ background level is reduced 
by tightening the accepted $\Delta{E}$ range or 
by applying more restrictive KID requirements; the continuum
background is reduced by tighter requirements on $LR$ and $r$. 
The effect of the $\Delta t$ tail is checked by tightening the $\Delta t$ range.
We do not observe any systematic variation in the fit results
when the  $\Delta E$, KID, $LR$, $r$, and $\Delta t$ requirements
are changed, as shown in Table~\ref{tbl:cutdep}.
To account for a possible $\Delta E$ tail, we repeat the fit with an additional
Gaussian function in the $\Delta E$ shape of
the $\pi^+\pi^-$ signal and the $K^+\pi^-$ background
. The fit yields
$\apipif = +0.75$ and $\spipif = -1.21$, consistent with our main results.
In addition, we divide the data into the 42~fb$^{-1}$ sample used for
our previous measurement and the recently added sample of
36~fb$^{-1}$. The result of the new analysis on the first 42~fb$^{-1}$ sample
is consistent with the published result~\cite{Acp_pipi_Belle}, and with
that for the more recent 36~fb$^{-1}$ sample.
%%%%%%%%%%%%%%%%%%%%%%%%%%%%%%%%%%%%%%%%%%%%%%%%%%%%%%%%%%%%%%%%%%%%%%%
\begin{table}
\caption{Selection-requirement dependence of $\apipif$ and $\spipif$ 
(MINOS errors only).}
\begin{ruledtabular}
\begin{tabular}{c|cc} 
Cut value   & $\apipif$               & $\spipif$ \\ \hline
default      & $0.77^{+0.20}_{-0.23}$ & $-1.23^{+0.24}_{-0.15}$ \\
(KID $< 0.4$)   &   &  \\ \hline
$|\Delta E| < 2\sigma$ & $0.81^{+0.20}_{-0.22}$ &  $-1.21^{+0.25}_{-0.16}$ \\
$|\Delta E| < 1\sigma$ & $0.82^{+0.21}_{-0.25}$ &  $-1.18^{+0.29}_{-0.19}$ \\ \hline
KID $< 0.20$ & $0.74^{+0.20}_{-0.23}$ & $-1.11^{+0.26}_{-0.17}$ \\
KID $< 0.15$ & $0.59^{+0.22}_{-0.24}$ & $-1.14^{+0.23}_{-0.14}$ \\ \hline
$LR > 0.825$ & $0.84^{+0.22}_{-0.25}$ & $-1.19^{+0.27}_{-0.18}$ \\ 
$LR > 0.925$ & $0.69^{+0.26}_{-0.30}$ & $-1.24^{+0.30}_{-0.19}$ \\ \hline
$|qr| > 0.75$  & $1.02^{+0.19}_{-0.25}$ & $-1.24^{+0.19}_{-0.25}$ \\
$|qr| > 0.875$ & $0.91^{+0.24}_{-0.31}$ & $-1.18^{+0.24}_{-0.31}$ \\ \hline
$|\Delta t| < 15$~ps & $0.77^{+0.20}_{-0.23}$ & $-1.25^{+0.24}_{-0.15}$ \\
$|\Delta t| < 5$~ps  & $0.76^{+0.20}_{-0.22}$ & $-1.27^{+0.26}_{-0.17}$ \\ \hline
Sample~I (42~fb$^{-1}$) & $1.00^{+0.19}_{-0.25}$ & $-1.14^{+0.30}_{-0.21}$ \\
Sample~II (36~fb$^{-1}$) & $0.37^{+0.32}_{-0.33}$ & $-1.99^{+0.70}_{-0.65}$ \\
\end{tabular}
\end{ruledtabular}
\label{tbl:cutdep}
\end{table}

A comparison of the event yields and $\Delta t$ distributions for
$B^0$- and $\bzb$-tagged events in the sideband region reveals
no significant asymmetry as shown in Fig.~\ref{fig:dt-control}(a).
We also use samples of non-$CP$ eigenstate $B^0 \to D^-\pi^+$,
$D^{*-}\pi^+$ and $D^{*-}\rho^+$ decays, selected with
the same event-shape criteria, to check for biases in the analysis.
The combined fit to this control sample of 15321 events yields
${\cal A} = -0.015 \pm 0.022$ 
and
${\cal S} = 0.045 \pm 0.033$.  
The $\Delta t$ distribution for this sample is shown in
Fig.~\ref{fig:dt-control}(b).
As expected,
neither mixing-induced nor direct $CP$-violating asymmetry is observed.
%%%%%%%%%%%%%%%%%%%%%%%%%%%%%%%%%%%%%%%%%%%%%%%%%%%%%%%%%%%%%%%%%%%%%%%%%
\begin{figure}[!htbp]
\resizebox{0.6\textwidth}{!}{\includegraphics{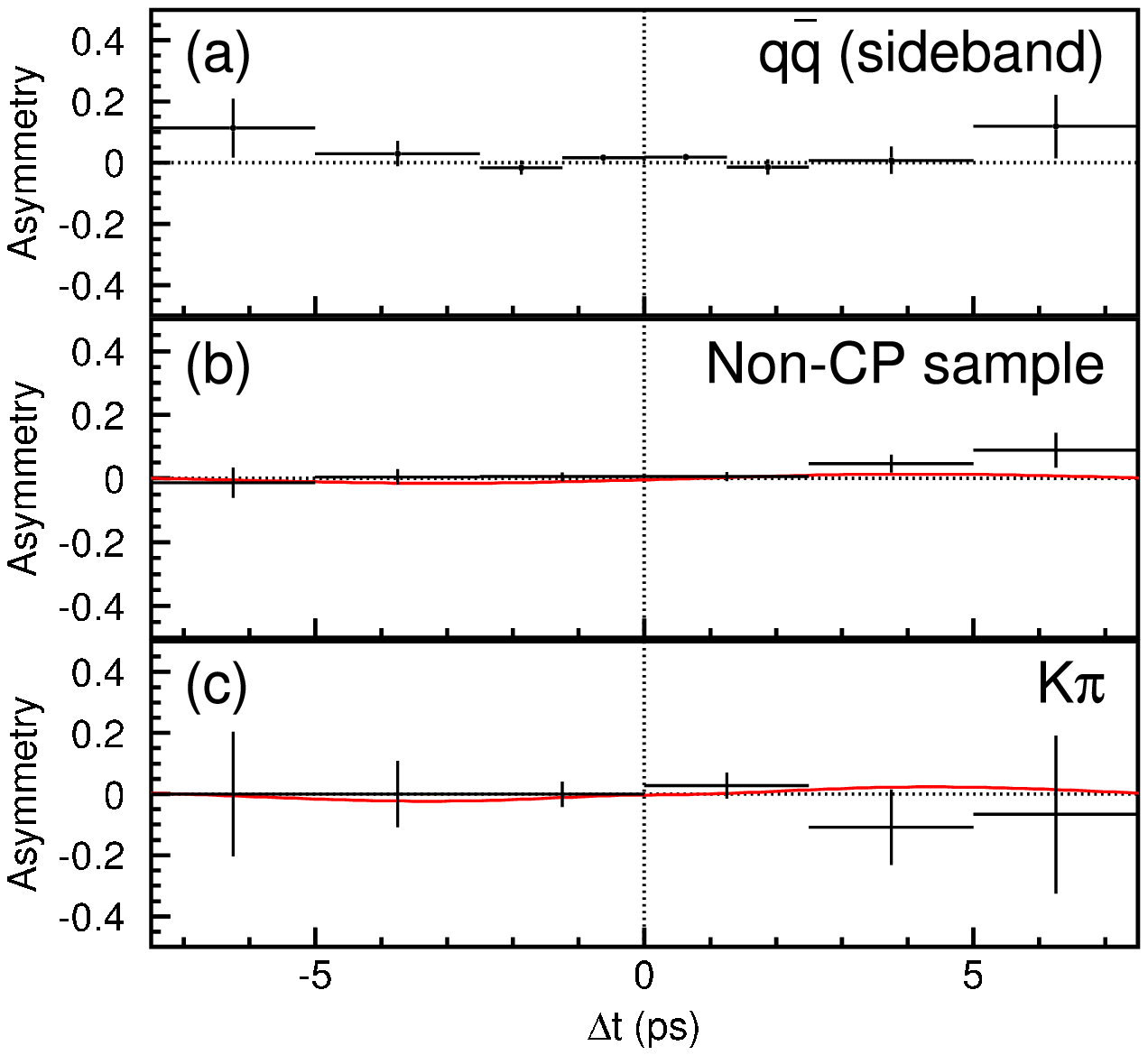}}
\caption{
The distributions of the raw $\Delta t$ asymmetries 
for (a) $B^0 \to \pi^+\pi^-$ sideband events,
(b) the $B^0 \to D^-\pi^+$, $D^{*-}\pi^+$ and 
$D^{*-}\rho^+$ candidates combined and (c) $B^0 \to K^+\pi^-$ candidates.
Fit curves are also shown.}
\label{fig:dt-control}
\end{figure}
%%%%%%%%%%%%%%%%%%%%%%%%%%%%%%%%%%%%%%%%%%%%%%%%%%%%%%%%%%%%%%%%%%%%%%%%%

We select $B^0 \to K^+\pi^-$ candidates
by positively identifying the charged kaons.
A fit to the 1371 
candidates (610 signal events)
yields $\akpif = -0.03 \pm 0.11$, in agreement with the
counting analysis mentioned above~\cite{dcpv_ichep02},
and $\skpif = 0.08 \pm 0.16$, which is consistent with zero as shown in
Fig.~\ref{fig:dt-control}(c).
The MINOS errors for ${\akpi}$ and ${\skpi}$ are consistent with those
from MC pseudo-experiment models of the $B^0 \rightarrow K^+\pi^-$
measurement as shown in Fig.~\ref{fig:toymc_kpi_err}.
%%%%%%%%%%%%%%%%%%%%%%%%%%%%%%%%%%%%%%%%%%%%%%%%%%%%%%%%%%%%%%%%%%%%%%%%%%%
\begin{figure}[!htbp]
\begin{center}
\resizebox{0.6\textwidth}{!}{\includegraphics{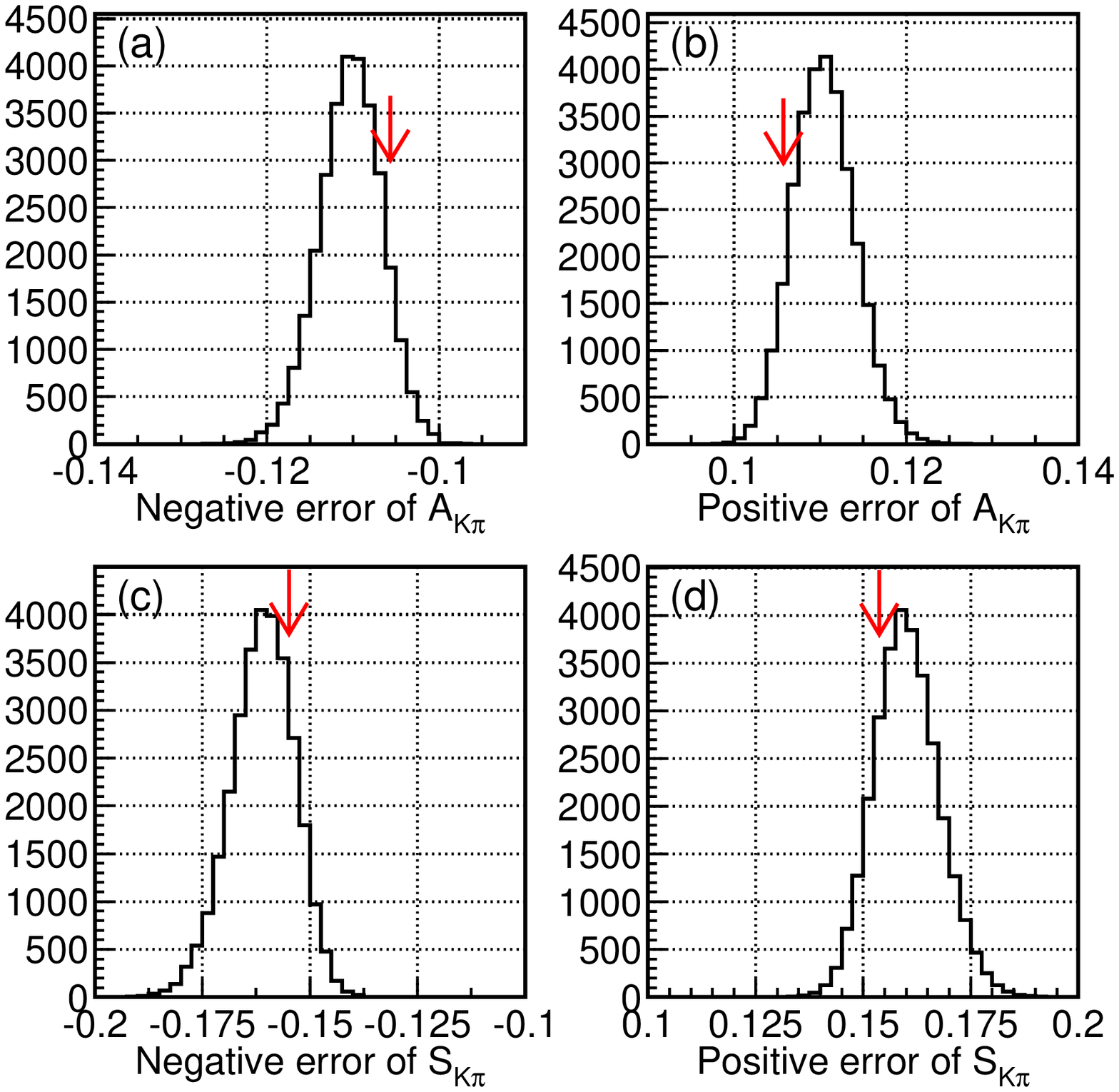}}
\caption{The result of MC pseudo-experiments for $B^0$ $\rightarrow$ $K^+\pi^-$
with input values of $\akpi = -0.03$ and $\skpi = 0.08$: 
the distributions of (a) the negative and (b) positive MINOS errors of $\akpi$, and
(c) the negative and (d) positive MINOS errors of $\skpi$.
The arrows indicate the MINOS errors obtained from the fit to data.} 
\label{fig:toymc_kpi_err}
\end{center}
\end{figure}
%%%%%%%%%%%%%%%%%%%%%%%%%%%%%%%%%%%%%%%%%%%%%%%%%%%%%%%%%%%%%%%%%%%%%%%%%
With the $K^+\pi^-$ event sample, we use the vertex
reconstruction method and wrong-tag fractions described
in Secs.~\ref{sec:FLTAG} and \ref{sec:vertex}  and determine
$\taub = 1.46\pm 0.08$~ps [Fig.~\ref{fig:lifetime}(b)] and 
$\dmd = 0.55^{+0.05}_{-0.07}$~ps$^{-1}$, which are
in agreement with the world average values~\cite{PDG}.

The selected $K^+\pi^-$ sample and the kaon mis-identification
probability measured from a sample of inclusive $D^{*+}\to D^0(\to
K^-\pi^+)\pi^+$ and $\phi \to K^+K^-$ decays are used to make independent 
estimates of the $K^+\pi^-$ background fractions in the $\pi^+\pi^-$ sample.
The results are $32 \pm 2 ~K^+\pi^-$ events in the signal region
with $LR > 0.825$ and $15 \pm 2~K^+\pi^-$ events with $LR \leq 0.825$;
these values are consistent with the results of the fit used to  
determine $\spipif$ and $\apipif$.
The changes in
$\apipif$ ($^{+0.005}_{-0.0}$) and $\spipif$ ($^{+0.0}_{-0.03}$) when
these $K^+\pi^-$ background fractions are used are
included in the systematic error associated with the background fraction.
%%v4<<
%%v6 The effect of charge asymmetry of a misidentification rate from kaons,
%%v6 which is described in Sec.~\ref{sec:recb}, is negligibly small in
%%v6 $\apipi$ and $\spipi$.
The effect of a possible charge asymmetry in the kaon misidentification rate,
described in Sec.~\ref{sec:recb}, is negligibly small.
%%v4>>

We check the measurement of $\apipif$ using time-independent fits to 
the $\Delta E$ distributions for the $B^0$ and $\overline{B}^0$ tags.
We determine the yields from fits to the $\Delta E$ distributions for
each of the 12 $LR$-$r$ bins for 
the $B^0$ and $\overline{B}^0$ tags separately (24 fits in total). 
We obtain $\apipif = 0.56^{+0.26}_{-0.27}$, which is consistent with 
the time-dependent $CP$ fit result.

 As discussed above, the nominal fit result is outside of the physical region. 
We also consider fits that constrain
the results to be in the physical region defined by
$\apipif^2+\spipif^2\le 1$.
The disadvantage of the constrained fitting method is that 
when the fit result is close to the physical boundary,
the errors returned from the fit are not Gaussian and are difficult to
interpret.
A constrained fit finds
$\apipif=\avaluecnst$ and $\spipif=\svaluecnst$, 
on the boundary of the physical region;
$\chi^2$ values that are defined
in Sec.~\ref{sec:result} for the $\Delta t$ projections are
$\chi^2$ = 12.4/12~DOF (13.6/12~DOF) for the $B^0$ ($\overline{B}^0$) tag.

\subsection{Significance}
\label{sec:CL}
We use the Feldman-Cousins frequentist approach~\cite{FeldmanCousins}
to determine the statistical significance of our measurement.
In order to form confidence intervals, we use
the $\apipi$ and $\spipi$ distributions of the results of fits to
MC pseudo-experiments for various input values of $\apipi$ and $\spipi$.
The distributions incorporate possible biases
at the boundary of the physical region as well as
a correlation between $\apipif$ and $\spipif$;
these effects are taken into account by this method.
The distributions are also smeared with Gaussian functions that
account for systematic errors.
The details of the method used to obtain the confidence intervals are described 
in Appendix~\ref{app:ToyMC}.
Figure~\ref{fig:2dcl} shows the resulting two-dimensional confidence regions 
in the $\apipi$ vs. $\spipi$ plane.
The case that $CP$ symmetry is conserved, $\apipi=\spipi=0$, is ruled out
at the $\cldd$ confidence level (C.L.), equivalent to $3.4\sigma$ significance for Gaussian errors.
The minimum confidence level for $\apipi$ = 0, the case of no direct $CP$ violation, occurs 
at $(\spipi, \apipi$) = ($-1.0, 0.0)$ and is 97.3$\%$,
which corresponds to 2.2$\sigma$ significance.

If the source of $CP$ violation is only due to $B-\overline{B}$ mixing
or $\Delta B=2$ transitions as in so-called 
superweak scenarios~\cite{bib:Bigi,bib:Wolfenstein},
then ($\spipi, \apipi$) = ($-{\rm sin}2\phi_1, 0$).
The C.L. at this point is 98.1$\%$, equivalent to 2.3$\sigma$ significance.

%%%%%%%%%%%%%%%%%%%%%%%%%%%%%%%%%%%%%%%%%%%%%%%%%%%%
\begin{figure}[!htb]
\resizebox{0.6\textwidth}{!}{\includegraphics{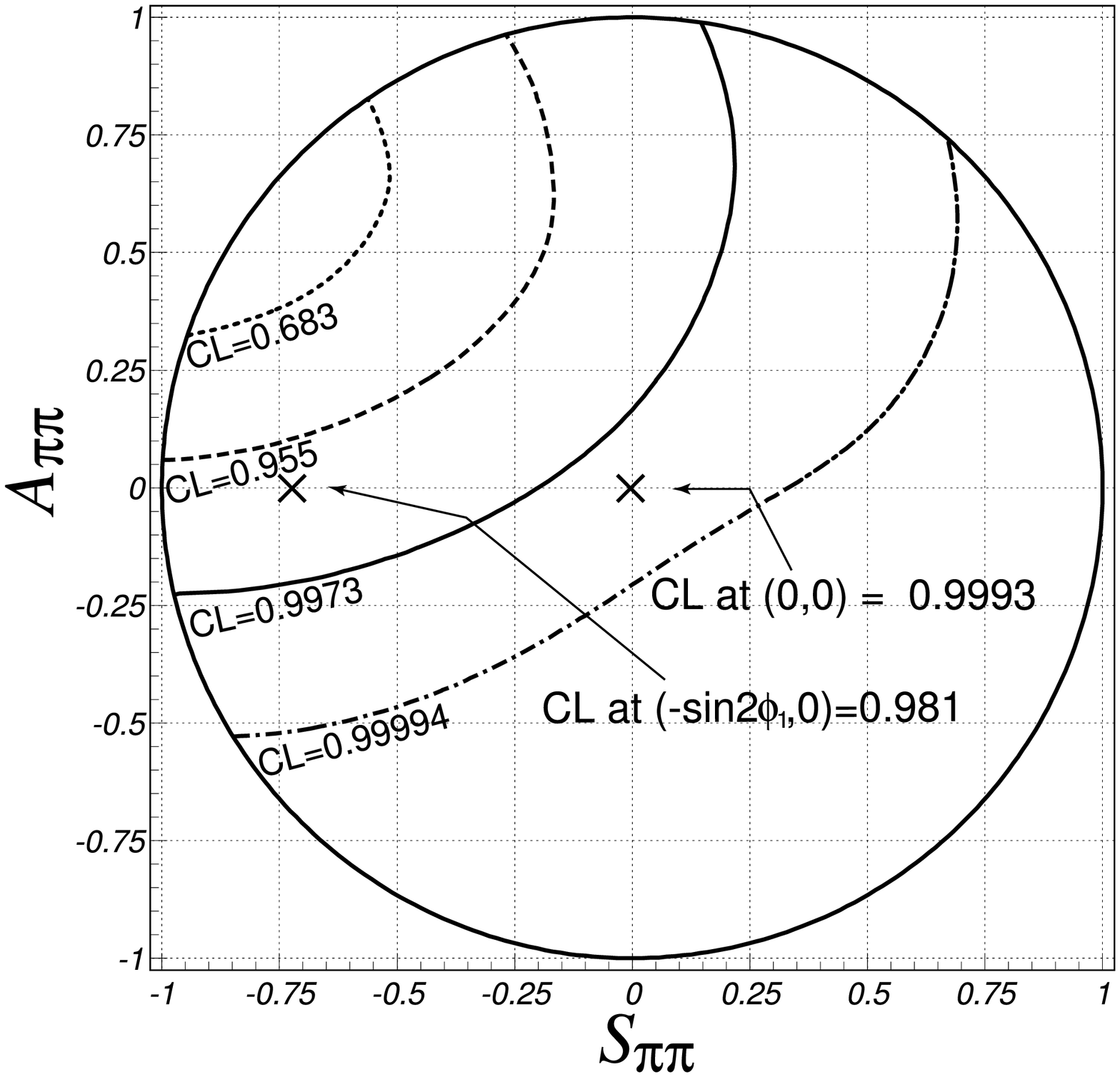}}
\caption{Confidence regions for $\apipi$ and $\spipi$.}
\label{fig:2dcl}
\end{figure}
%%%%%%%%%%%%%%%%%%%%%%%%%%%%%%%%%%%%%%%%%%%%%%%%%%%

\section{Discussion}
\label{sec:phi2}

Using the standard definitions of weak phases $\phi_1$, $\phi_2$, and $\phi_3$,
the decay amplitudes for $B^0$ and $\overline{B}^0$ to $\pi^+\pi^-$ are
\begin{eqnarray}
A(B^0\rightarrow\pi^+\pi^-) &=& -(|T|e^{i\delta_{T}}e^{i\phi_3}~~ +
|P|e^{i\delta_P}) , \nonumber \\
A(\overline{B}^0\rightarrow\pi^+\pi^-) &=& -(|T|e^{i\delta_{T}}e^{-i\phi_3} +
|P|e^{i\delta_P}) ,
\end{eqnarray}
where $T$ and $P$ are the amplitudes for the tree and penguin 
graphs and $\delta_T$ and $\delta_P$ are their strong phases. 
Here we  adopt the notation of Ref.~\cite{bib:SA_TH_GR} and use
the convention in which the top-quark contributions are integrated out
in the short-distance effective Hamiltonian. In addition, the unitarity relation
$V^*_{ub}V_{ud}$ + $V^*_{cb}V_{cd}$ = $-V^*_{tb}V_{td}$ is applied. 
Using the above expressions and $\phi_2$ = $\pi - \phi_1 - \phi_3$, we determine
\begin{eqnarray}
\lambda_{\pi\pi}\equiv 
e^{2i\phi_2}\frac{1+|P/T|e^{i(\delta+\phi_3)}}
{1+|P/T|e^{i(\delta-\phi_3)}} ~~.
\end{eqnarray}
Explicit expressions for $\spipi$ and $\apipi$ are
\begin{eqnarray}
\spipi &=& [{\rm sin}2\phi_2 + 2|P/T|{\rm sin}(\phi_1 - \phi_2){\rm cos}{\delta} \nonumber \\
& & \mbox{} - |P/T|^2{\rm sin}2\phi_1]/\rpipi,  \nonumber \\
\apipi &=& - [2|P/T|{\rm sin}(\phi_2 + \phi_1){\rm sin}{\delta}]/\rpipi, \nonumber \\
\rpipi &=& 1 - 2|P/T|{\rm cos}{\delta}{\rm cos}(\phi_2 + \phi_1) + |P/T|^2 ,
\end{eqnarray}
where $\delta$ $\equiv$ $\delta_P$ $-$ $\delta_T$.
We take $-180^\circ\leq\delta\leq{180}^\circ$.
When $\apipi$ is positive and $0^\circ<\phi_1+\phi_2<180^\circ$, $\delta$ is 
negative.

Recent theoretical estimates 
prefer $|P/T| \sim 0.3$ with large uncertainties~\cite{bib:GR,bib:LR,bib:Beneke,bib:phi2-th}. 
Figures~\ref{fig:phi2-delta}(a)-(e) show the regions for $\phi_2$ and $\delta$  
corresponding to the 68.3$\%$ C.L., 95.5$\%$ C.L. and 99.73$\%$ C.L. region of
 $\apipi$ and $\spipi$  
(shown in Fig.~\ref{fig:2dcl}) for representative values of $|P/T|$ and $\phi_1$.  
Note that a value of ($\spipif$,$\apipif$) inside the 68.3$\%$ C.L. contour
requires a value of $|P/T|$ greater than $\sim$0.3.

The allowed region is not very sensitive to variations
of $\phi_1$ within the errors of the measurements,
as can be seen by comparing Figs.~\ref{fig:phi2-delta}(a), (c) and (e).
The range of $\phi_2$ that corresponds to the 95.5$\%$ C.L. region of 
$\apipi$ and $\spipi$ in Fig.~\ref{fig:2dcl} is
\begin{eqnarray*}
78^\circ \leq \phi_2 \leq 152^\circ,
\end{eqnarray*}
for $\phi_1=23.5^\circ$ and $0.15\leq|P/T|\leq0.45$.
The result is in agreement with constraints on the unitarity triangle
from other measurements~\cite{bib:NIR02}.
%%%%%%%%%%%%%%%%%%%%%%%%%%%%%%%%%%%%%%%%%%%%%%%%%%%%%%%%%%%%%%%%%%%%%%%%
\begin{figure*}[!htbp]
\resizebox{0.90\textwidth}{!}{\includegraphics{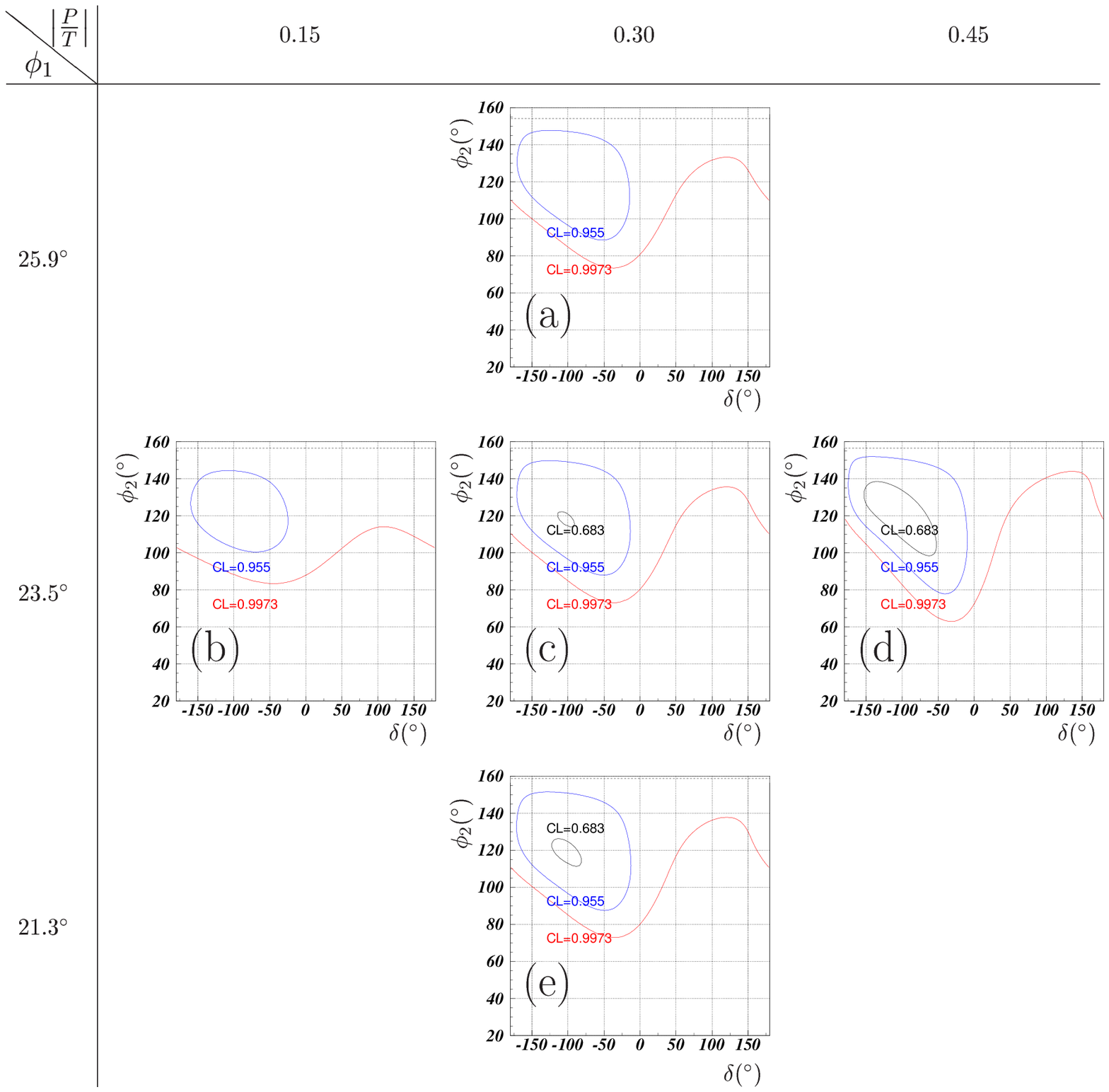}}
\caption{
The regions for $\phi_2$ and $\delta$ corresponding to the 68.3$\%$, 95.5$\%$, and 
99.73$\%$ C.L. regions of $\apipi$ and $\spipi$ in Fig.~\ref{fig:2dcl} for
(a) $\phi_1=25.9^\circ$, $|P/T|$=0.3, (b) $\phi_1=23.5^\circ$, $|P/T|$=0.15,
(c) $\phi_1=23.5^\circ$, $|P/T|$=0.3, (d) $\phi_1=23.5^\circ$, $|P/T|$=0.45,
and (e) $\phi_1=21.3^\circ$, $|P/T|$=0.3. The horizontal dashed lines correspond to $\phi_2 = 180^\circ - \phi_1$.}
\label{fig:phi2-delta}
\end{figure*}
%%%%%%%%%%%%%%%%%%%%%%%%%%%%%%%%%%%%%%%%%%%%%%%%%%%%%%%%%%%%%%%%%%%%%%%

\section{CONCLUSION}
\label{sec:conclusion}
In summary, we have performed an improved measurement of
$CP$ violation parameters in $B^0 \rightarrow \pi^+\pi^-$ decays.
An unbinned maximum likelihood fit to 
760 $B^0$ $\rightarrow$ $\pi^+\pi^-$ candidates,
which contain $163^{+24}_{-23}$(stat) $\pi^+\pi^-$ signal events,
yields $\apipif = \aresult$, and $\spipif = \sresult$,
where the statistical uncertainties are determined from MC pseudo-experiments.
This result is consistent with our previous measurement~\cite{Acp_pipi_Belle}
and supersedes it.
We obtain confidence intervals
for $CP$-violating asymmetry parameters $\apipi$ and $\spipi$
based on the Feldman-Cousins approach 
where we use MC pseudo-experiments to determine acceptance regions.
We rule out the $CP$-conserving case, $\apipi=\spipi=0$, at 
the $\cldd$ confidence level. 

The result for $\spipi$ indicates that mixing-induced $CP$ violation
is large, and the large $\apipi$ term is an indication of direct
$CP$ violation in $B$ meson decay. 
Constraints within the Standard Model
on the CKM angle $\phi_2$ and the hadronic phase
difference between the tree ($T$) and penguin ($P$) amplitudes are obtained for
$|P/T|$ values that are favored theoretically.
We find an allowed region of $\phi_2$ that is consistent with
constraints on the unitarity triangle from other measurements.

\section*{ACKNOWLEDGMENTS}
We wish to thank the KEKB accelerator group for the excellent
operation of the KEKB accelerator.
We acknowledge support from the Ministry of Education,
Culture, Sports, Science, and Technology of Japan
and the Japan Society for the Promotion of Science;
the Australian Research Council
and the Australian Department of Industry, Science and Resources;
the National Science Foundation of China under contract No.~10175071;
the Department of Science and Technology of India;
the BK21 program of the Ministry of Education of Korea
and the CHEP SRC program of the Korea Science and Engineering Foundation;
the Polish State Committee for Scientific Research
under contract No.~2P03B 17017;
the Ministry of Science and Technology of the Russian Federation;
the Ministry of Education, Science and Sport of the Republic of Slovenia;
the National Science Council and the Ministry of Education of Taiwan;
and the U.S.\ Department of Energy.

%%%%%%%%%%%%%%%%%%%%%%%
%%%%%%%%%%%%%%%%%%%%%%%
\appendix
\section{MC pseudo-experiments and confidence regions}
\label{app:ToyMC}
We use ensembles of Monte Carlo (MC) pseudo-experiments
to determine the significance of our measurement
and obtain confidence regions.
They are also used for various crosschecks.
Each pseudo-experiment consists of events that are generated
with the nominal PDFs, which are incorporated in Eq.~\ref{eq:likelihood}.
Since the parameters in the PDFs are derived from 
large-statistics control samples and sideband events, 
the pseudo-experiments precisely reproduce $\Delta t$ distributions
that are consistent with data.
In particular, they are free from possible discrepancies between
data and GEANT-based detector simulation.

To generate each event in a pseudo-experiment, 
we first choose one $LR$-$r$ region $m$ 
randomly from a population that is based on
the regional event fractions obtained from data.
We then generate $\Delta E$ and $\MBC$ values with
distributions that are determined by
the event fractions $g^m_{\pi\pi}$, $g^m_{K\pi}$ and
$g^m_{q\overline{q}}$, which are listed in Table~\ref{tbl:frac}.
The values of the probability functions $f^m_{\pi\pi}$, $f^m_{K\pi}$ and 
$f^m_{q\overline{q}}$ (in Eq.~\ref{eq:likelihood}) are determined from the
$\Delta E$ and $\MBC$ values.
We randomly choose an event type, $\pi\pi$, $K\pi$, $q\overline{q}$
or outlier, from a population
based on $f^m_{\pi\pi}$, $f^m_{K\pi}$,
$f^m_{q\overline{q}}$, and the outlier fraction $f_{ol}$.
We generate $q$, $\Delta t$ and resolution parameters
according to the PDF of the selected event type.

We repeat this procedure until the number of events reaches
the observed number of events (760 events), and
perform an unbinned maximum likelihood fit
to obtain $\xapipi$ and $\xspipi$, which
are the fit results and should be distinguished from
the true (input) values $\apipi$ and $\spipi$.
To account for the systematic error, 
each fit result is further modified by an amount
determined from a Gaussian variation.
We test the entire procedure using GEANT simulation, and find that 
distributions of $\xapipi$ and $\xspipi$ obtained from
the GEANT experiments are in good agreement with those from
pseudo-experiments, when the resolution functions in the PDF are
extracted from a lifetime fit to the GEANT data.
We also verify that there is no fit bias as shown in Fig.~\ref{fig:linearity}.
%%%%%%%%%%%%%%%%%%%%%%%%%%%%%%%%%%%%%%%%%%%%%%%%%%%
\begin{figure}[!htbp]
\begin{center}
\resizebox{0.4\textwidth}{!}{\includegraphics{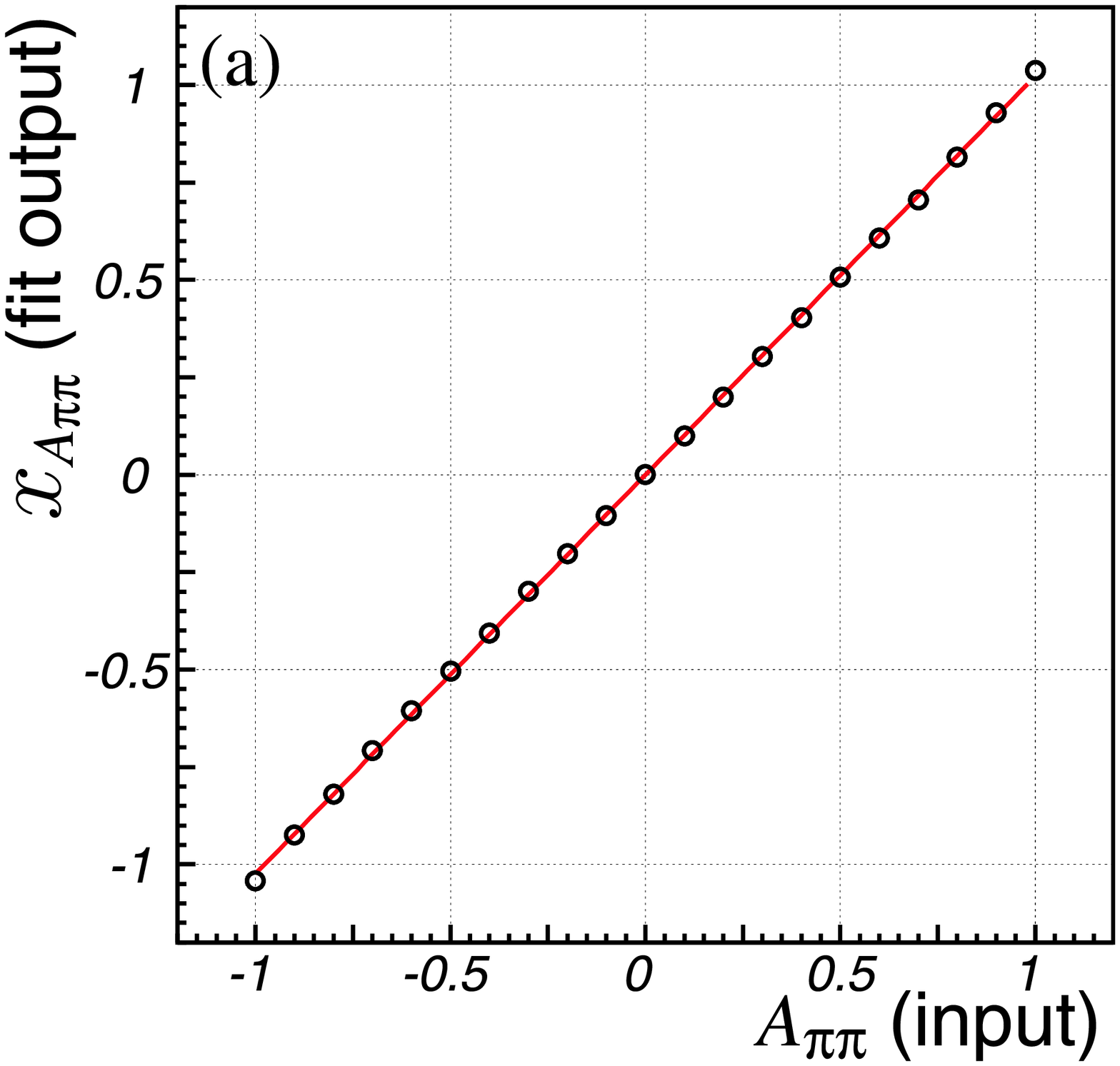}}
\resizebox{0.4\textwidth}{!}{\includegraphics{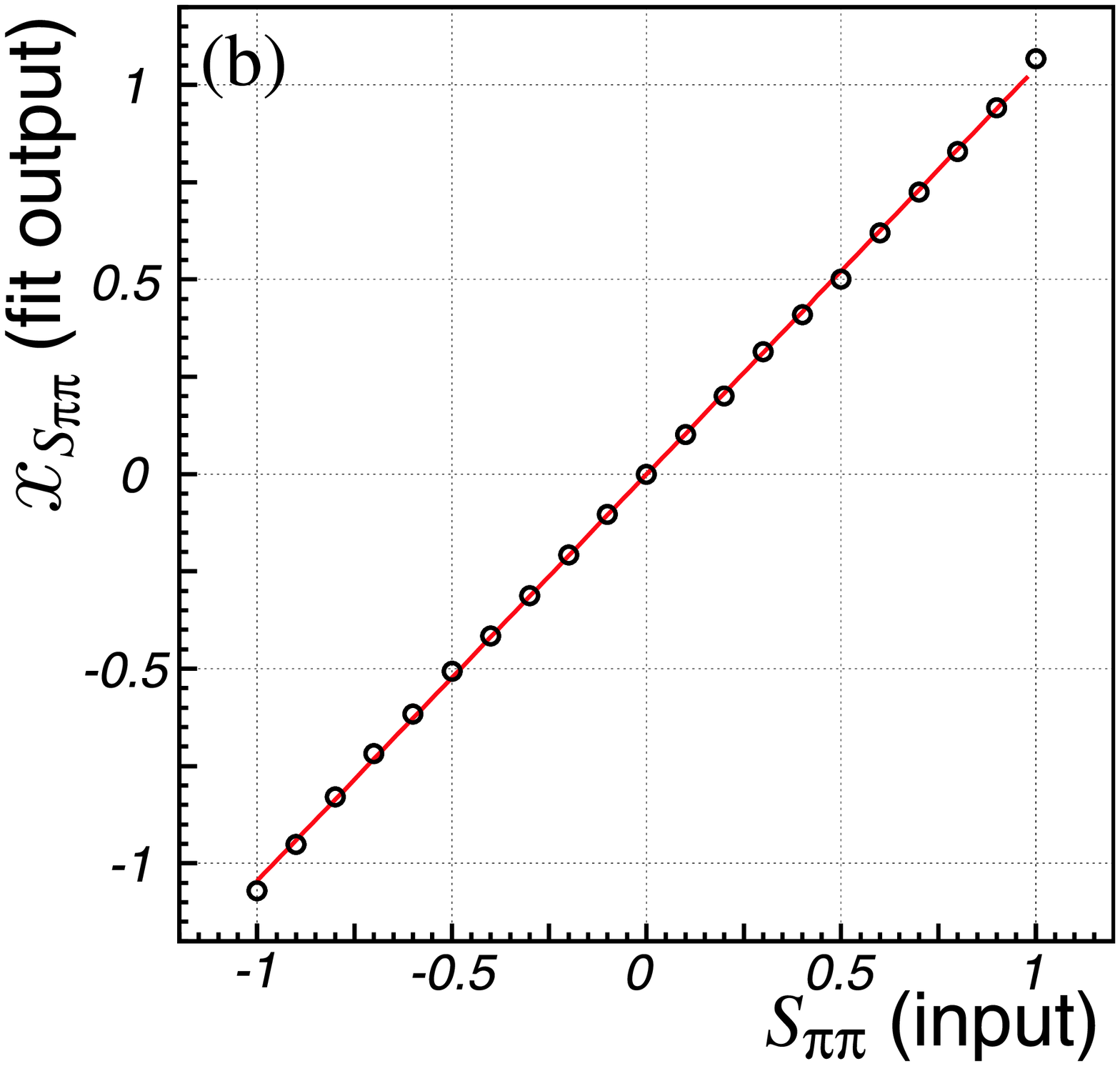}}
\caption{
Mean values of fit results vs. input values of MC
pseudo-experiments for (a) $\apipif$ and (b) $\spipif$.
The solid lines are linear fit results.
}
\label{fig:linearity}
\end{center}
\end{figure}
%%%%%%%%%%%%%%%%%%%%%%%%%%%%%%%%%%%%%%%%%%%%%%%%%%%%

We adopt the Feldman-Cousins frequentist approach~\cite{FeldmanCousins},
which is based on the likelihood-ratio ordering principle,
to obtain the confidence regions that are shown
in Fig.~\ref{fig:2dcl}~\cite{bib:Nakadaira}. 
In the following, we first illusrate how we can obtain 1-dimensional confidence
intervals for $\apipi$ with $\spipi$ set to zero; intervals for $\spipi$ are obtained in a very similar way.
We then explain the method used for the determination of 
the two-dimensional confidence regions for $\apipi$ and $\spipi$, which is
an extension of that for the 1-dimensional case.

We generate 10,000 experiments for 317 sets of ($\apipi$, $\spipi$)
values that cover the entire physical region.
The fit to each set of experiments yields an $\xapipi$ distribution
that depends on the input $\apipi$ value. To account for this dependence,
we use a PDF for $\xapipi$ that consists of two Gaussian functions
whose parameters depend on $\apipi$:
\begin{eqnarray*}
P(\xapipi|\apipi) = f_A \cdot G(\xapipi;m_1,\sigma_1) \nonumber \\
+ (1-f_A)\cdot G(\xapipi;m_2,\sigma_2),
\end{eqnarray*}
where $G(x;m,\sigma)$ 
represents a Gaussian function with mean $m$ and standard deviation $\sigma$, and
$f_A$, $m_{1(2)}$, and $\sigma_{1(2)}$ are polynomials of $\apipi$.
The explicit expressions for $f_A$, $m_1$, $m_2$, $\sigma_1$ 
and $\sigma_2$ are
\begin{eqnarray*}
f_A &=& a_1 + a_2{\apipi}^2, \\
m_1 &=& a_3 + a_4\apipi, \\
\sigma_1 &=& a_5 + a_6{\apipi}^2, \\
m_2 &=& a_7 + a_8\apipi + a_9{\apipi}^2 + a_{10}{\apipi}^3, \\
\sigma_2 &=& a_{11} + a_{12}\apipi,
\end{eqnarray*}
where the 12 free parameters $(a_i$, $i=1,12$) are determined from
an unbinned maximum-likelihood fit to the
$\xapipi$ distributions.
Figures~\ref{fig:xapipi}(a) and (b) show the distributions and
the $\xapipi$ PDF
for the cases $(\apipi,\spipi) = (0,0)$ and 
$(\apipi,\spipi) = (1,0)$, respectively.
The PDFs are in good agreement with the distributions of pseudo-experiments in both cases.
%%%%%%%%%%%%%%%%%%%%%%%%%%%%%%%%%%%%%%%%%%%%%%%%%%%%%%%%%%%%%%%%%%%%%%%%%%%%%%%%%
\begin{figure}[!htb]
\begin{center}
\resizebox{0.4\textwidth}{!}{\includegraphics{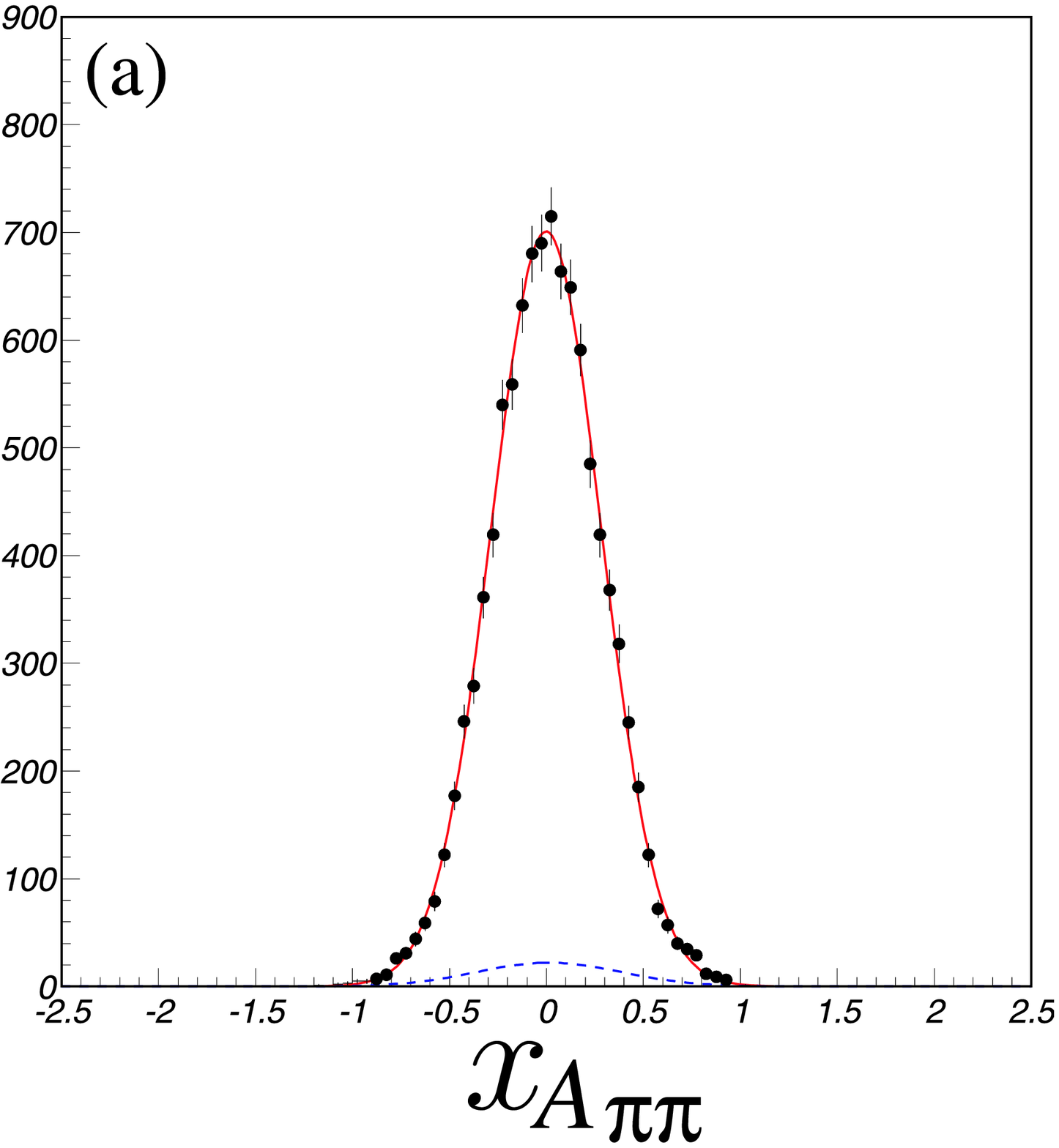}}
\resizebox{0.4\textwidth}{!}{\includegraphics{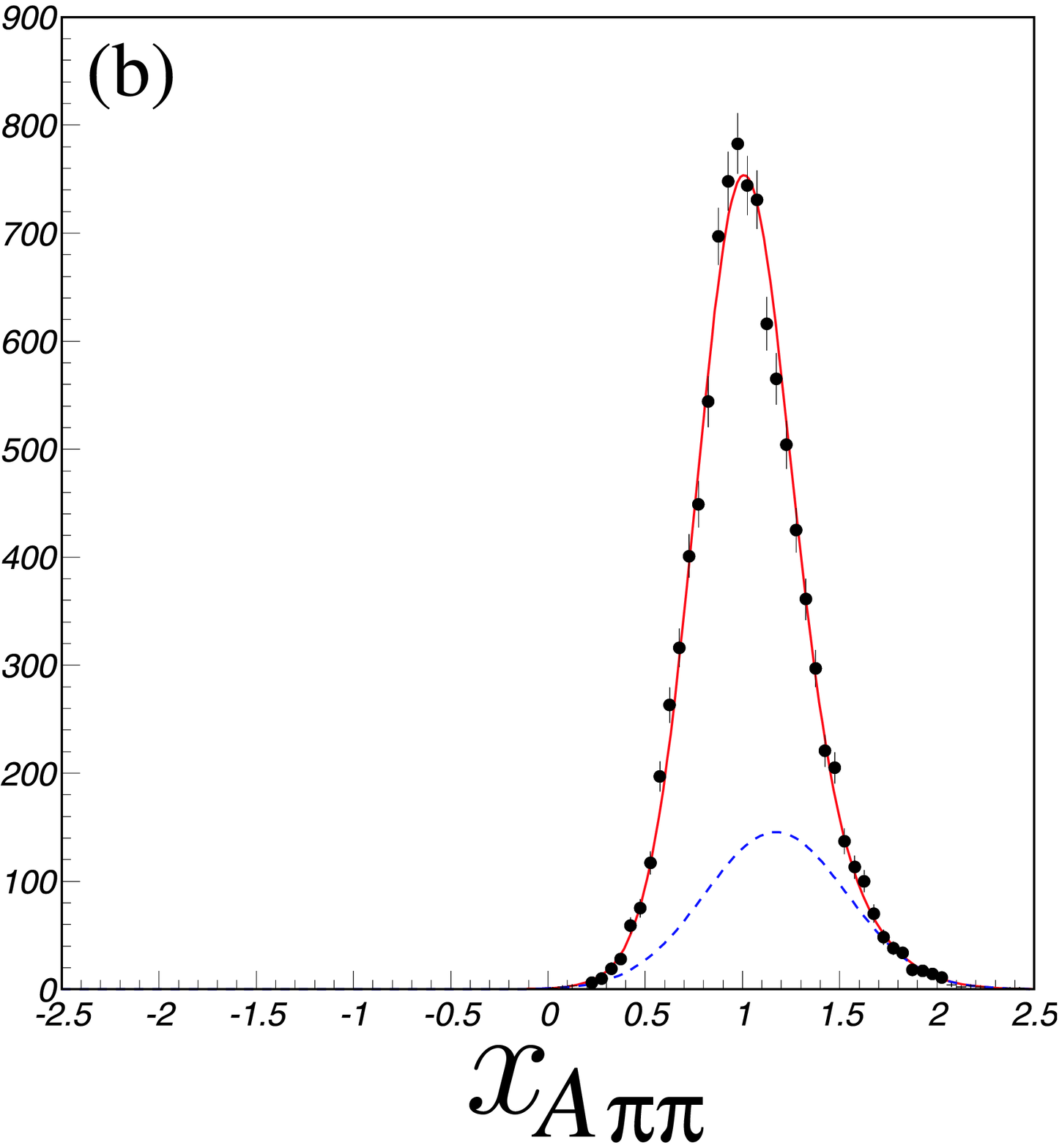}}
\end{center}
\caption{$\xapipi$ distributions and the PDFs for (a) $(\apipi,\spipi)=(0,0)$ and
(b) $(\apipi,\spipi)=(+1,0)$. Solid and dashed curves represent the total PDFs and
the second Gaussian components, respectively.}
\label{fig:xapipi}
\end{figure}
%%%%%%%%%%%%%%%%%%%%%%%%%%%%%%%%%%%%%%%%%%%%%%%%%%%%%%%%%%%%%%%%%%%%%%%%%%%%%%%%

The acceptance region $[\xamin, \xamax]$ 
for a given $\apipi$ and a confidence level $\alpha$ is defined by:
\begin{eqnarray*}
\alpha = \int_{\xamin}^{\xamax}d\xapipi P(\xapipi|\apipi).
\end{eqnarray*}
We adopt the likelihood-ratio ordering principle to
determine $\xamin$ and $\xamax$. 
Using the likelihood-ratio
\begin{eqnarray*}
LR(\xapipi|\apipi) \equiv P(\xapipi|\apipi)/P(\xapipi|\abest),
\end{eqnarray*}
where $\abest$ gives the maximum $P$ value for a given $\xapipi$,
we require 
\begin{eqnarray*}
LR(\xapipi|\apipi) \ge LR(\xamin|\apipi)=LR(\xamax|\apipi)
\end{eqnarray*}
for any $\xapipi$ in $[\xamin, \xamax]$.
Figure~\ref{fig:cbelt} shows the resulting confidence belts for $\apipi$.
%%%%%%%%%%%%%%%%%%%%%%%%%%%%%%%%%%%%%%%%%%%%%%%%%%%%%%%%%%%%%%%%%%%%%%%%%%%%%%
\begin{figure}[!htb]
\begin{center}
\resizebox{0.6\textwidth}{!}{\includegraphics{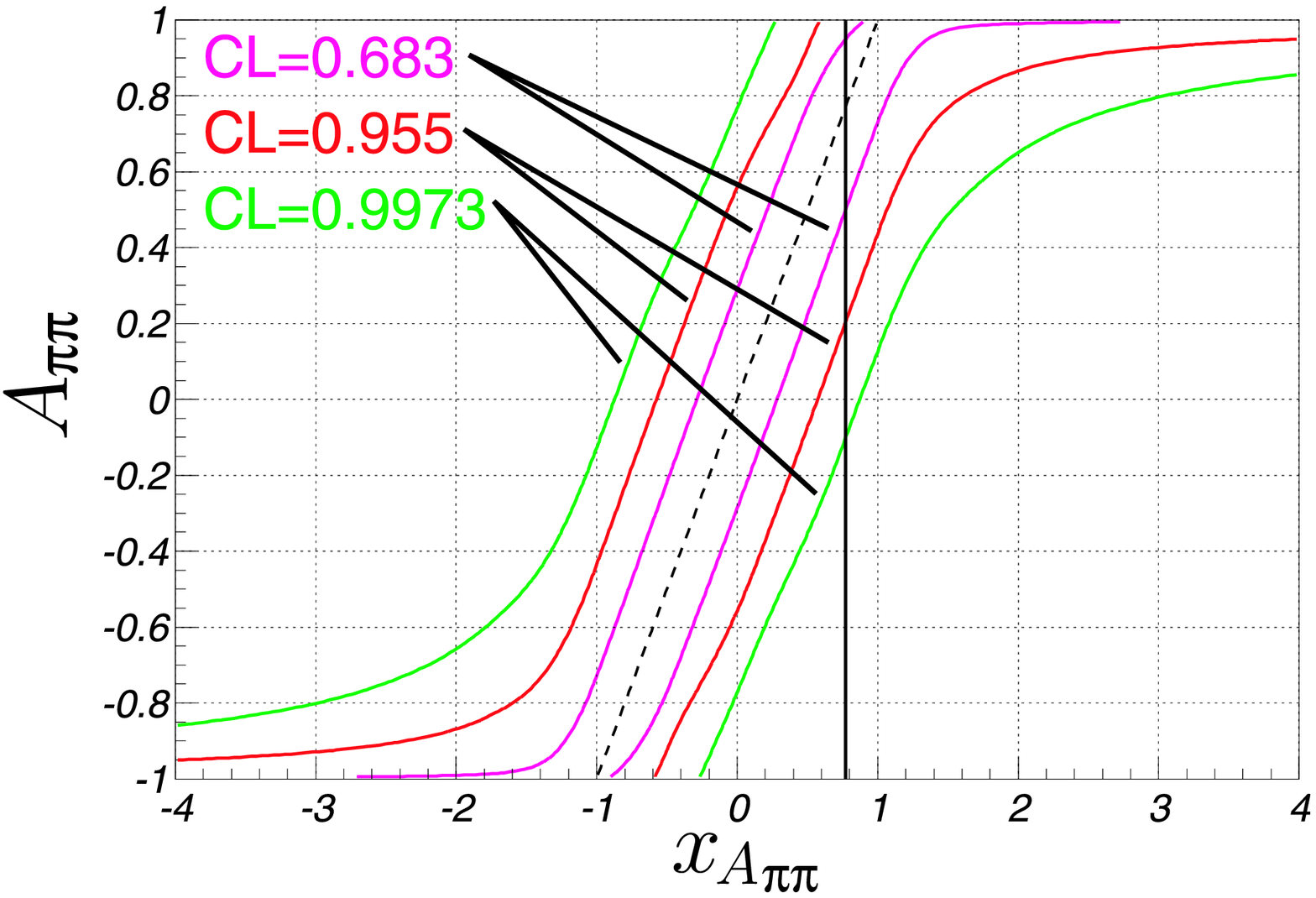}}
\end{center}
\caption{Confidence belts in the $\apipi$ versus $\xapipi$ plane
for $\alpha =$ 0.683, 0.955, and 0.9973  in the one-dimensional case.
The dashed line corresponds to $\apipi = x_{{\cal A}\pi\pi}$.}
\label{fig:cbelt}
\end{figure}
%%%%%%%%%%%%%%%%%%%%%%%%%%%%%%%%%%%%%%%%%%%%%%%%%%%%%%%%%%%%%%%%%%%%%%%%%%%%%
For a given measurement $\xapipi$,
a confidence interval at a confidence level $\alpha$ is obtained from the figure.

The procedure to obtain the 2-dimensional confidence regions for $\apipi$ and $\spipi$
(Fig.~\ref{fig:2dcl}) is an extension of the method described above.
We use the following PDF:
\begin{eqnarray*}
&P(&\xapipi,\xspipi|\apipi,\spipi~) \\
 &=& f_{AS}\cdot G(\apipi;m_{A1},\sigma_{A1})
                           \cdot G(\spipi;m_{S1},\sigma_{S1}) \nonumber \\
 &+& (1-f_{AS})\cdot G(\apipi;m_{A2},\sigma_{A2})
            \cdot G(\spipi;m_{S2},\sigma_{S2}),
\end{eqnarray*}
where $f_{AS}$, $m_{A1(2)}$, $m_{S1(2)}$, $\sigma_{A1(2)}$ and
$\sigma_{S1(2)}$ depend both on $\apipi$ and $\spipi$.
There are 27 free parameters that are determined from
an unbinned maximum-likelihood fit to the ($\xapipi$, $\xspipi$) distributions.
We find that the PDFs represent the distributions of $\xapipi$ and $\xspipi$
very well for the input $\apipi$ and $\spipi$ values
over the entire physical region.
An acceptance region $\Omega$ at a confidence level $\alpha$ is also defined
in a similar way to that for the 1-dimensional case:
\begin{eqnarray*}
\alpha = \int_\Omega d\xapipi d\xspipi P(\xapipi,\xspipi|\apipi,\spipi),
\end{eqnarray*}
where likelihood-ratio ordering is used. Using the requirement
\begin{eqnarray*}
LR(\xapipi,\xspipi|\apipi,\spipi) \ge LR(+0.77,-1.23|\apipi,\spipi)
\end{eqnarray*}
which corresponds to an acceptance region 
with our measurement $(\xapipi,\xspipi)=(+0.77,-1.23)$ at its boundary,
we scan the physical region in the $\apipi$-$\spipi$ plane and 
calculate a confidence level $\alpha$ for each 
input point $(\apipi,\spipi)$ to obtain the confidence regions shown in Fig.~\ref{fig:2dcl}.

\section{Source of small MINOS errors}
\label{app:small-error}
The Feldman-Cousins approach with acceptance regions
determined from MC pseudo-experiments,
which is described in Appendix~\ref{app:ToyMC}, is
applicable to a wide range of analyses.
On the other hand, care is needed when using experimental MINOS errors for
the confidence interval calculation, as mentioned in Sec.~\ref{sec:stat_error}.
In particular, difficulties
may arise when the number of events is not large and
the true values of physical parameters are located
close to a physical boundary.
In such a case, a small number of events can have a large influence 
on both the size of the MINOS errors and the shape of the log-likelihood ratio curve.
The likelihood function for some events may become negative
when the fit parameters are beyond the physical boundary.

The observed features of the MINOS errors arise
when there is an event that
restricts the fit parameters in or close to the physical region,
while the fit to all the other events gives a maximum likelihood
that is located outside the physical region and
is not allowed by the aforementioned restrictive event.
For example, in this fit the removal of such a restrictive event
results in an $\spipif$ value that is more negative than $\spipif = -1.23$
(further from the physical boundary).
In this case, 
the log-likelihood ratio curve is deformed by inclusion of the restrictive event,
even if the curve before the inclusion is well-described by a parabola.
The sizes of the MINOS errors also become small.

We investigate 
this type of single-event fluctuation and its relation to
the size of the MINOS errors with MC pseudo-experiments.
For each experiment, we repeat the fit
by removing each event in turn. The event that creates
the largest difference in $\spipif$ is tagged as the
restrictive event and the change produced by the removal of the
restrictive event, $\dspipif$, is recorded.
When we choose the point of maximum likelihood at the physical boundary
($\apipi$, $\spipi$) = ($\avaluecnst$, $\svaluecnst$)
as the input for the MC pseudo-experiments,
we obtain the average values of
$\dspipif$ as a function of the positive error of $\spipif$
shown in Fig.~\ref{fig:ses}.
The correlation between the size of the error and
the single-event fluctuation is evident.
%%%%%%%%%%%%%%%%%%%%%%%%%%%%%%%%%%%%%%%%%%%%%%%%%%%%
\begin{figure}[!htb]
\resizebox{0.6\textwidth}{!}{\includegraphics{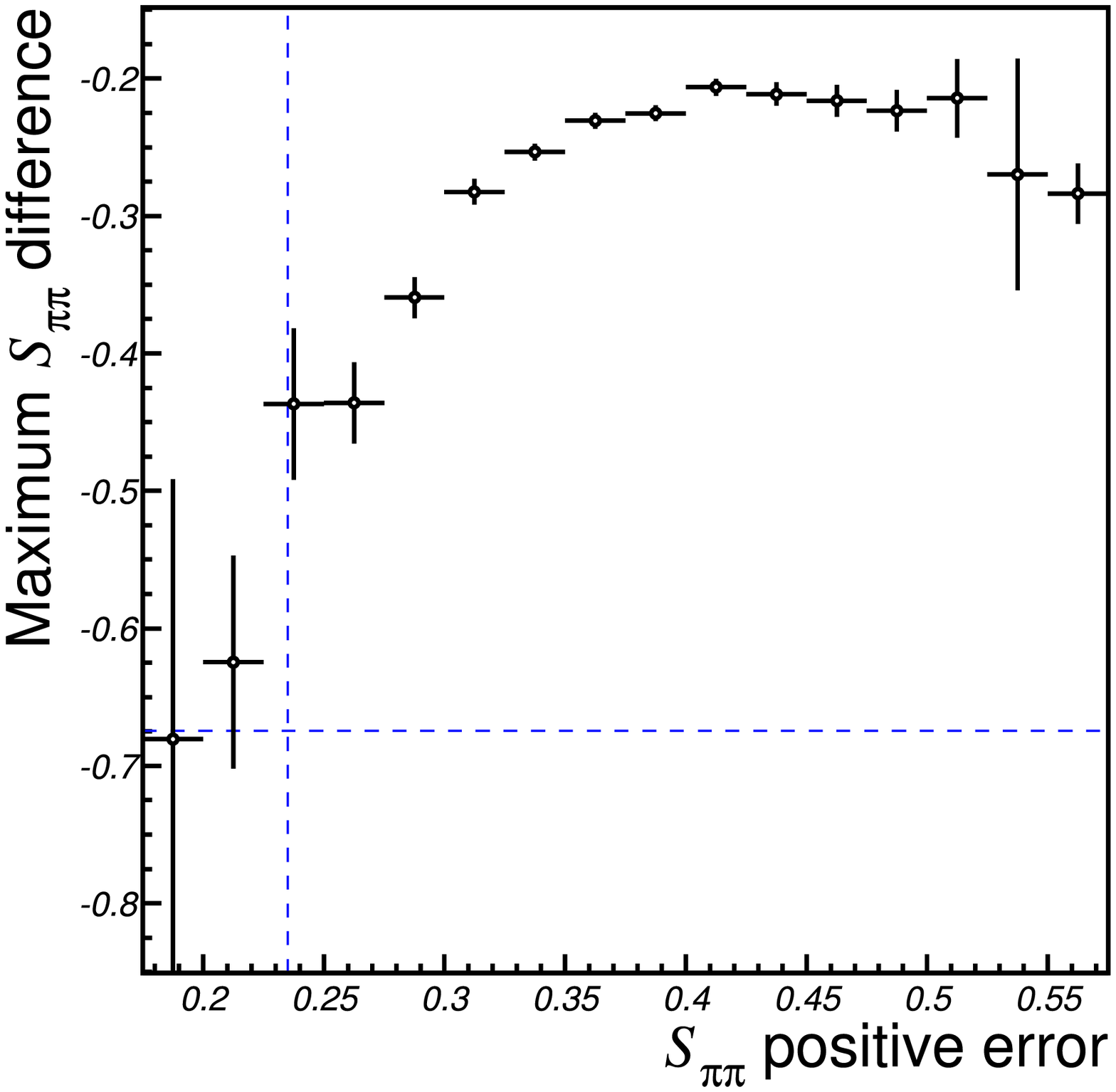}}
\caption{Single-event fluctuation versus the positive MINOS error on $\spipif$.
The dashed lines indicate the observed $\dspipif$ and $\spipif$ positive MINOS error.}
\label{fig:ses}
\end{figure}
%%%%%%%%%%%%%%%%%%%%%%%%%%%%%%%%%%%%%%%%%%%%%%%%%%%

In our data, we have
one event that has a large effect on the sizes of MINOS errors.
The removal of this event from the fit gives
$\spipif=-1.91^{+0.36}_{-0.33}$ and $\apipif=0.64^{+0.19}_{-0.20}$,
where the errors are MINOS errors;
$\spipif$ is shifted to a more
negative value ($\dspipif = -0.67$) and the MINOS error increases.
This event has $qr=-0.92$ which is 
close to unambiguous $B$ flavor assignment
and corresponds to a very small wrong-tag probability.
In addition, this event has $\Delta E=-0.01$~GeV, and $LR=0.98$,  
which corresponds to small $B^0 \rightarrow K^+\pi^-$ 
and $q\bar{q}$ background probabilities.
For this event, $\Delta t=-3.8$~ps and, thus,
sin($\Delta{m_d}\Delta{t}$)$\approx{-1}$. According to Eq.~\ref{eq:likelihood},
this event has a negative likelihood value at negative $\spipif$ values beyond
$\sim -1.5$, where it truncates the log-likelihood ratio curve.
As a result, the negative MINOS error for the entire event sample is 
restricted by this single event.

As shown in Fig.~\ref{fig:ses},
the observed single-event fluctuation $\dspipif = -0.67$
is consistent with the expectation from the MC pseudo-experiments
if the positive error of $\spipif$ is $\sim +0.24$,
which is the case for our data.
A similar study for input values of $\spipi$ and $\apipi$
that are well within the physically allowed region indicates that
this behavior occurs much less often.


\begin{thebibliography}{99}
\bibitem{CC}
Throughout this paper, 
the inclusion of the charge conjugate mode decay is implied
unless otherwise stated.
\bibitem{footnoteStatError}
We obtain MINOS errors $^{+0.20}_{-0.23}$ for $\apipif$ and $^{+0.24}_{-0.15}$ for $\spipif$ 
for data, while the MC pseudo-experiments yield
$^{+0.24}_{-0.26}$ for $\apipif$ and $^{+0.36}_{-0.32}$ for $\spipif$
with the input values at the boundary and
$\pm$0.27 for $\apipif$ and $\pm$0.37 for $\spipif$ at $\apipi$ = $\spipi$ = 0.
The correlation between $\apipif$ and $\spipif$ is found to be 0.024 for data.
\bibitem{bib:phi1}
Using the average of recent values of sin2$\phi_1$ 
from Belle~\cite{CP1_Belle3} and BaBar~\cite{CP1_BaBar},
we obtain $\phi_1=(23.5^{+2.4}_{-2.2})^\circ$.
\end{thebibliography}
\end{document}